\documentclass[showpacs,preprintnumbers,superscriptaddress,amsmath,amssymb,twocolumn,floatfix]{revtex4}
\usepackage{graphicx}
\usepackage{dcolumn}
\usepackage{bm}
\usepackage{bbm}
\usepackage{psfrag}
\usepackage{epsfig}
\usepackage{amsmath}
\usepackage{amssymb}
\usepackage{color}
\usepackage{mathbbol}
\usepackage{subfigure}
\usepackage{epstopdf,txfonts,hyperref}

\newcommand{\rp}[1]{(\ref{#1})}

\newcommand{\abs}[1]{\left|{#1}\right|}

\newcommand{\av}[1]{\left\langle #1 \right\rangle}

\newcommand{\al}[1]{^{(#1)}}
\newcommand{\da}{^\dagger}

\newcommand{\pt}[1]{\left( #1 \right)}
\newcommand{\pq}[1]{\left[ #1 \right]}
\newcommand{\pg}[1]{\left\{ #1 \right\}}

\newcommand{\lpq}[1]{\left[ #1 \right.}
\newcommand{\lpg}[1]{\left\{ #1 \right.}

\newcommand{\rpq}[1]{\left. #1 \right]}
\newcommand{\rpg}[1]{\left. #1 \right\}}
\newcommand{\ee}{{\rm e}}
\newcommand{\ii}{{\rm i}}
\newcommand{\dd}{{\rm d}}

\newcommand{\id}{\mathbbm{1}}

\newcommand{\nn}{{\nonumber}}

\newcommand{\va}{{\bf a}}

\newcommand{\AAA}{{\cal A}}

\newcommand{\GG}{{\cal G}}

\newcommand{\QQ}{{\cal Q}}

\newcommand{\WW}{{\cal W}}

\newcommand{\ZZ}{{\cal Z}}

\begin{document}

\title{Stationary entanglement of photons and atoms in a high-finesse resonator}

\author{Hessam Habibian}
\affiliation{Theoretische Physik, Universit\"{a}t des Saarlandes, D-66123 Saarbr\"{u}cken, Germany}
\affiliation{Grup d'\'{O}ptica, Departament de F\'{i}sica, Universitat Aut\`{o}noma de Barcelona, E-08193 Bellaterra, Barcelona, Spain}
\affiliation{ICFO Ð Institut de Ci\'encies Fot\'oniques, Mediterranean Technology Park, E-08860 Castelldefels (Barcelona), Spain}

\author{Stefano Zippilli}
\affiliation{Dipartimento di Ingegneria Industriale, Universit\`a degli Studi di Salerno, Via Giovanni Paolo II 132, I-84084 Fisciano (SA), Italy}

\author{Fabrizio Illuminati}
\affiliation{Dipartimento di Ingegneria Industriale, Universit\`a degli Studi di Salerno, Via Giovanni Paolo II 132, I-84084 Fisciano (SA), Italy}
\affiliation{INFN, Istituto Nazionale di Fisica Nucleare, Sezione di Napoli Gruppo collegato di Salerno, I-84084 Fisciano (SA), Italy}
\affiliation{CNISM, Consorzio Interuniversitario per la Fisica della Materia, Unit\`a di Salerno, I-84084 Fisciano (SA), Italy}

\author{Giovanna Morigi}
\affiliation{Theoretische Physik, Universit\"{a}t des Saarlandes, D-66123 Saarbr\"{u}cken, Germany}

\date{\today}

\begin{abstract}
We predict that the collective excitations of an atomic array become entangled with the light of a high-finesse cavity mode when they are suitably coupled. This entanglement is of Einstein-Podolsky-Rosen type, it is robust against cavity losses and is a stationary property of the coupled system. It is generated when the atomic array is aligned along the cavity axis and driven transversally by a laser, when coherent scattering of photons into the cavity mode is suppressed because of phase-mismatching. We identify the parameter regimes under which entanglement is found and show that these are compatible with existing experimental setups.
\end{abstract}


\maketitle

\section{Introduction}

Entanglement is a quantum mechanical feature with no classical counterpart and is a central resource for emerging and future quantum technologies~\cite{Horodecki}. Entangled states were first introduced in the seminal paper by Einstein, Podolsky, and Rosen in the context of nonlocality of quantum mechanics~\cite{EPR}. In a continuous-variable setting, Einstein-Podolsky-Rosen (EPR) type of states are intimately connected with two-mode squeezing~\cite{Reid1989,Reid2009}, to the extent that the amount of noise reduction below the shot-noise limit can be used as an entanglement measure~\cite{Duan,Adesso}. Due to the relative simplicity
in their generation and manipulation compared to entangled states in systems of discrete variables, continuous-variable entangled states are very attractive and play a key role in several protocols for quantum sensing~\cite{QSensing}, quantum metrology~\cite{Monras,QMetrology}, and quantum information processing and secure telecommunication~\cite{vanLoock,Hammerer,Lee,Weedbrook,Krauter2013}.

\begin{figure}
\centering
\includegraphics[width=6.5cm]{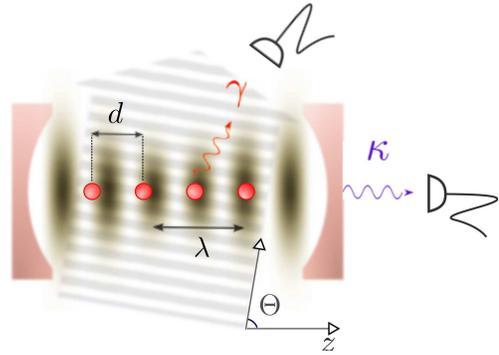}
\caption[]{(Color online) An array of two-level atoms at inter-particle distance $d$ is confined along the $z$-axis of a standing-wave optical cavity of wavelength $\lambda$, which couples to the atoms with strength $g$ and decays at rate $\kappa$. The atoms are also driven by a laser with Rabi frequency $\Omega$, whose wave vector forms an angle $\Theta$ with the cavity axis. The properties of the light scattered into the resonator are determined, amongst others, by the ratio $d/\lambda$. This configuration can be realized by confining the atoms in a deep optical lattice, see Refs.~\cite{Brakhane,Schleier-Schmidt}. }
\label{Fig:model}
\end{figure}

In the present work we analyze the generation of entanglement between the collective excitations of an atomic array and the mode of a high-finesse optical resonator, to which the atoms of the array couple. The setup we consider is sketched in Fig.~\ref{Fig:model} and is similar to the one realized experimentally in Ref.~\cite{Brakhane,Schleier-Schmidt}. We make use of the phase-matching properties due to the periodic order of the atomic medium in order to select the nonlinear optical process giving rise to non-degenerate parametric amplification~\cite{Ou1992}, whose entanglement can be tuned through the intensity and frequency of the laser that coherently drives the atoms. We show that this dynamics gives rise to EPR-type of states between dipolar collective excitations and the cavity field, whose entanglement is robust against cavity decay and spontaneous emission and is realized at the steady state of the interaction dynamics. This protocol can be instrumental for implementing continuous-variable quantum interfaces with atomic ensembles in optical resonators~\cite{ParkinsKimble}, and provides an alternative to existing schemes based on hot atomic ensembles in free space~\cite{Hammerer}.

This work is organized as follows. In Sec.~\ref{Sec:2} we introduce and discuss the model; in Sec.~\ref{Sec:Gauss} we reformulate the problem in terms of the covariance matrix, from which one can completely characterize the system dynamics in the Gaussian limit. In Sec.~\ref{Results} we present and discuss results both for the transient dynamics and at steady state. Conclusions and outlook for future research are discussed in Sec.~\ref{Conclusions}, while the appendix reports the mathematical details of
the analysis discussed in Sec.~\ref{Sec:Gauss}.

\section{Theoretical Model}
\label{Sec:2}

The system that we consider is illustrated in Fig~\ref{Fig:model}: $N$ atoms are pinned along the axis of a high-finesse optical resonator and form a one-dimensional periodic array with inter-particle distance (lattice spacing) $d$. Their optical dipole transition couples with a single mode of the resonator at wavelength $\lambda$ and is also driven by an external laser field, whose wave vector forms an angle $\Theta$ with the cavity axis. The coherence properties of the light scattered by the atoms into the cavity is controlled by the ratio $d/\lambda$.This system is equal to the one discussed in~\cite{HabibianPRA2011}. However, here, we are interested in 
parameter regimes not investigated in~\cite{HabibianPRA2011}. 

In this section we report the main equations of the system dynamics referring the reader to Ref.~\cite{HabibianPRA2011} for a detailed analysis of their derivation. 

\subsection{Hamiltonian}

The relevant degrees of freedom are the ones of the cavity mode harmonic oscillator and the electronic excitations of the atoms, which are (quasi)-resonantly coupled by the cavity and the laser, while the laser is here a classical plane wave with wave vector $k$ and frequency $\omega_p$. The motion of the atoms can be neglected by assuming that they are pinned at the minima of an optical lattice which is parallel to the cavity axis and has periodicity $d$.

By appropriately choosing the polarization of the fields, the initial state of the atoms, and possibly the re-pumping lasers, only a two-level electronic transition is effectively driven by the fields. In this case, the internal atomic levels are labeled by the ground state $|1\rangle_j$ and the excited state $|2\rangle_j$, forming a dipolar transition at frequency $\omega_0$, with $j=1,\ldots,N$ labeling the atoms. We introduce the pseudo-spin operators $S_j=|1\rangle_j\langle2|$ and $S_j^z=\frac{1}{2}(|2\rangle_j\langle2|-|1\rangle_j\langle1|)$, with $[S_j,S_\ell^{\dagger}]=-2\delta_{j,\ell}S_j^z$. The cavity mode is at frequency $\omega_c$ and wave number $k=2\pi/\lambda$. Treating it in second quantization, we denote by $a$, $a^{\dagger}$ the associated annihilation and creation operators, respectively, with canonical commutation relation $[a,a^{\dagger}]=1$. In a reference frame rotating at the laser frequency $\omega_p$ the Hamiltonian governing the coherent dynamics coupling the cavity and the spins reads
\begin{align}\label{fullHamil}
\mathcal{H}=&\hbar\delta_c a^{\dag}a+\hbar\omega_z\sum_{j=1}^{N}
S_j^z+\sum_{j=1}^{N}\hbar g_j \left(S_j^{\dag}a+{\rm H.c.}\right)\nn\\
&+\ii\hbar\sum_{j=1}^{N}\left(\Omega_jS_j^{\dag}-{\rm H.c.}\right)\,,
\end{align}
where $\delta_c=\omega_c-\omega_p$  ($\omega_z=\omega_0-\omega_p$) is the detuning between cavity (atomic transition) and laser frequency.

The position ${\bf r}_j$ of the atoms  within the spatial-mode function of cavity and laser fields determines the interaction  strengths $g_j$ and $\Omega_j$. For ${\bf r}_j=(0,0,z_j)$, with $z_j=jd$, then
\begin{eqnarray}
&&g_j=g\cos(kz_j+\phi)\,\\
&&\Omega_j=\Omega\,e^{{\rm i}kz_j\cos\Theta}\,,
\end{eqnarray}
with $g$ the vacuum Rabi frequency, $\phi$ the phase mismatch between the atomic array and the cavity spatial mode function, $\Omega$ the laser Rabi frequency, and $\Theta$ the angle between the laser wave vector and the array axis. The wave vectors of laser and cavity field are here assumed to be equal, as their possible small difference has no effect on the results of our analysis.

Depending on the geometry of the setup, the array can scatter photons coherently into the cavity mode when the von Laue (Bragg) condition is satisfied. In the presence of a sufficiently strong coupling (verified when the total cooperativity is larger than unity) this behavior can lead to saturation of the intra-cavity field due to interference~\cite{Zippilli2004}. Further interesting regimes are accessed when the geometry is so chosen, that coherent scattering into the cavity mode is suppressed. In this case the cavity field is solely pumped by inelastic processes; then, nonclassical light can be generated on demand by matching the reciprocal lattice vector $G_0=2\pi/d$ of the array with a higher-order scattering process~\cite{Vogel,FernandezVidal,HabibianPRA2011}.

\subsection{Low-saturation limit}

We now focus on the low saturation limit, and assume that the laser Rabi frequency  is sufficiently weak so that $\Omega\ll \abs{\omega_z+\ii\, \gamma/2}$, where $\gamma$ is the atomic line width. In this limit the atoms are weakly excited, $\av{S_j\da S_j}\ll 1$, and the pseudo-spin operators can be expressed in terms of bosonic operators $b_j$ and $b_j\da$ via the Holstein-Primakoff transformation~\cite{HolsteinPrimakoff}, here reported at first order in the saturation parameter $s=\Omega/\abs{\omega_z+\ii\, \gamma/2}$:
\begin{align}
& S_j\approx \left(1-\frac{1}{2}b^\dag_jb_j\right)b_j\label{Sjbj}\,, \\
& S_j^z=b_j\da b_j+\frac{1}{2}\,,\label{Szbj}
\end{align}
with $\pq{b_j,b_\ell\da}=\delta_{j,\ell}$.

The relevant spin dynamics is conveniently studied in Fourier space for the collective spin-wave operators $$b_q=\frac{1}{\sqrt N}\sum_{j=1}^N b_j e^{-\ii q jd}\,,$$
where the wave number is defined in the Brillouin zone $q\in(-\frac{G_0}{2},\frac{G_0}{2}]$ with $G_0=2\pi/d$. It is convenient to write the wave numbers of cavity and laser in terms of quasi-momenta $Q$ and $Q'$ within the Brillouin zone, such that
\begin{eqnarray*}
&&Q=k+G\,,\\
&&Q'=k\cos\Theta +G'\,,
\end{eqnarray*}
with  $G, G'$ reciprocal lattice vectors \cite{HabibianPRA2011}. In the following, we will focus on the limit in which the von Laue condition is not satisfied, which is formally expressed by the inequality $Q\neq \pm Q'$. This implies that the atomic array does not elastically scatter photons into the cavity field. Therefore, the cavity mode can only be populated via inelastic, higher-order photon scattering processes. We will choose $Q'=G'$ for some reciprocal lattice vector $G'$, that is realized, for instance, when the laser wave vector is perpendicular to the cavity axis: $\Theta=\pi/2$.  We also assume that the atoms occupy the minima of an optical lattice with inter-particle distance $d=\lambda/2$, with $\lambda=2\pi/k$ the wavelength of the cavity mode. In this limit, at fourth order the modes which are coupled are the spin wave mode $Q$ (that couples with the cavity field $a$) and the spin-wave modes $Q$ and $Q'$ (which are mutually coupled by four-photon processes). One can thus reduce the Hilbert space to these three modes.  Following the procedure developed in Ref.~\cite{HabibianPRA2011}, the system Hamiltonian can be reduced to
\begin{equation}
\label{H:4order}
{\mathcal H}= {\mathcal H}^{(2)}+{\mathcal H}^{(4)}\,,
\end{equation}
where ${\mathcal H}^{(2)}$ is a quadratic term,
\begin{align}
&{\mathcal H}^{(2)}=\hbar\delta_c a^\dag a+\hbar\omega_z (b_Q^\dag b_Q+b_{Q'}^\dag b_{Q'})
\nn\\
&
+ \hbar \tilde g\sqrt{N}(b_Q^\dag a+{\rm H.c.})+i\hbar\Omega\sqrt{N}(b^\dag_{Q'}-{\rm H.c.})\,,
\label{H:2}
\end{align}
with $\tilde{g}=g\cos(\phi)$. The Hamiltonian term ${\mathcal H}^{(4)}$ describes the inelastic processes in lowest order, and reads
\begin{align}
& {\cal H}^{(4)}=\frac{-\hbar \tilde{g}}{2\sqrt{N}}\left(b_{Q'}^{\dagger}b_{Q'}^{\dagger} b_{Q}\,\delta_{Q',G/2}+2b_{Q'}^{\dagger} b_{Q}^{\dag} b_{Q'}+b_Q^\dag b_Q^\dag b_Q \right)a \nn\\
&-{\rm i}\frac{\hbar\Omega}{2\sqrt{N}}\left(2b_{Q'}^{\dagger}b_{Q}^{\dagger}b_{Q}+b_{Q}^{\dagger}b_{Q}^{\dagger}b_{Q'}\, \delta_{Q',G/2}\right) + {\rm H.c.}\,.
\end{align}
We remark that the coupling with other spin wave modes can be neglected by assuming that they are initially in the ground state. In the remainder of the paper we set $Q=G_0/2$ and $Q'=G_0$, namely, the atoms are at half-wavelength distance and the laser is perpendicular to the cavity axis.

\subsection{Effective Hamiltonian}

When the Hamiltonian is considered up to the quadratic term, Eq. \eqref{H:2}, only the spin-wave mode $Q'$ is pumped, while the cavity, if initially in the vacuum state, remains empty. Solving the Heisenberg equation of motion for $b_{Q'}$ at this order yields the equation
\begin{equation}
\label{b:Q}
b_{Q'}\approx-\ii\Omega\sqrt{N}e^{-i\phi_L}/\omega_z\,,
\end{equation}
which is here reported for $|\omega_z|\gg \gamma/2$. By substituting this relation into Eq. \eqref{H:4order} we obtain the effective Hamiltonian
\begin{align}\label{Heff}
{\mathcal H_{\rm eff}}&=\hbar\delta_c a^\dag a+\hbar\delta_b b_Q^\dag b_Q +\hbar\alpha_{\rm BS}(a^\dag b_{Q}+{\rm H.c.})\nn\\
&+\hbar\chi\left(b_Q^\dag b_Q^\dag b_Q a +{\rm H.c.}\right)+\hbar(\alpha_{Q} b_{Q}^{\dag2}+{\rm H.c.})
\nn\\
&+\hbar(\alpha_{Q,a} a^\dag b_{Q}^\dag+{\rm H.c.})\,,
\end{align}
whose detailed derivation is reported in Ref.~\cite{HabibianPRA2011}. Here, the parameter $\delta_b$, defined as
$$\delta_b=\omega_z+\frac{2\Omega^2}{\omega_z}\, ,$$
is the frequency of the spin in the reference frame of the laser, which includes the dynamical Stark shift due to the interaction with the laser light. The third term on the right-hand-side (rhs) of Eq.~\rp{Heff} describes the linear interaction between spins and cavity and has strength
$$\alpha_{\rm BS}=\tilde{g}\sqrt{N}\left(1-\frac{\Omega^2}{\omega_z^2}\right)\,,$$
where the second term is the correction due to the non-linear interaction with the spin mode $b_{Q'}$. The remaining three terms on the rhs of Eq.~\rp{Heff} originate solely from $\mathcal H^{(4)}$. The fourth term, with coefficient
\begin{equation}
\label{chi}
\chi =\frac{\tilde g}{2\sqrt N} \,,
\end{equation}
describes fourth-order processes involving the cavity field and the spin wave mode $b_Q$. Their strength is inversely proportional to $\sqrt{N}$, and their effect is negligible for a sufficiently large number of atoms. Finally, the last two terms  describe a parametric-amplifier type of dynamics and are nonvanishing only when the phase matching condition $Q'=G'/2$ holds. In detail, one term represents the squeezing of mode $b_Q$ and scales with the coefficient
\begin{eqnarray}
\alpha_{Q}=-\frac{\Omega^2 }{2\omega_z}\, ,
\end{eqnarray}
while the non-degenerate parametric amplification of the spin-wave $b_Q$ and of the cavity scales with the frequency
\begin{eqnarray}
\label{a:Qa}
\alpha_{Q,a}=\tilde{g}\sqrt{N}\frac{\Omega^2}{2\omega_z^2}\, .
\end{eqnarray}

\subsection{Incoherent dynamics}

We next consider the open-system dynamics that is realized as soon as one includes dissipative processes. Dissipation arises from: (i) decay through the cavity mirrors at rate $\kappa$ and (ii) spontaneous emission of the spin-wave mode $Q$ at rate $\gamma$. These processes do not couple spin-wave modes among themselves nor with the cavity field. In the reduced Hilbert space of the cavity mode and of the spin-wave mode, whose state is described by the density matrix $\varrho$, the system dynamics is governed by the master equation
\begin{equation}\label{ME}
 \partial_t{\varrho}=-\frac{\ii}{\hbar}[{\mathcal H}_{\rm eff},\varrho]+{\mathcal L}\varrho\,,
\end{equation}
where
\begin{align}
 {\mathcal L}\varrho=&\kappa(2\,a\varrho a^\dag -a^\dag a\varrho -\varrho a^\dag a)\nn\\
 &+\frac{\gamma}{2}(2\,b_Q\varrho b_Q^\dag -b_Q^\dag b_Q\varrho -\varrho b_Q^\dag b_Q)\,.
\label{Master:Eq}
\end{align}
Correspondingly, in the Heisenberg picture the Heisenberg-Langevin equations of motion for operators $a$ and $b_Q$ read
\begin{eqnarray}
\label{HL:a}
&&\dot{a}=\frac{\ii}{\hbar}[{\mathcal H}_{\rm eff},a]-\kappa a+ \sqrt{2\kappa}a_{\rm in}\,,\\
&&\dot{b}_Q=\frac{\ii}{\hbar}[{\mathcal H}_{\rm eff},b_Q]-\frac{\gamma}{2} b_Q+\sqrt{\gamma}b_{\rm in}(t)\,,
\label{HL:b}
\end{eqnarray}
where $ a_{\rm in}$ and $b_{\rm in}$ are the input-field operators, which are de-correlated from each other and have first and second moments $\av{a_{\rm in}(t)}=\av{b_{\rm in}(t)}=0$ and $\av{a_{\rm in}(t)a\da_{\rm in}(t')}=\av{b_{\rm in}(t)b\da_{\rm in}(t')}=\delta(t-t')$ (while here $\av{a\da_{\rm in}(t)a_{\rm in}(t')}=\av{b\da_{\rm in}(t)b_{\rm in}(t')}=0$).

The fields emitted by the cavity and by the atoms can be determined using the input-output formalism~\cite{MilburnWalls}, which relates the output field with the system and input-noise operators through the relations
\begin{eqnarray}
\label{aout}
&& a_{\rm out}(t)=\sqrt{2\kappa}\,a(t)-a_{\rm in}(t)\,,\nn\\
&& b_{\rm out}(t)=\sqrt{\gamma}\,b_Q(t)-b_{\rm in}(t)\,.
\end{eqnarray}

Here we consider the situation in which the detuning of the laser from the atomic transition is much larger than the atomic line width, i.e., $|\omega_z|\gg\gamma/2 $. In this limit, the use of the effective Hamiltonian~\eqref{Heff} in the master equation~\eqref{ME} is correct as long as the coherent rates $\alpha_{Q,a}$, $\alpha_Q$, and $\alpha_{BS}$ are of the same order or larger than the dissipation rates $\kappa,\gamma/2$~\cite{FernandezVidal,HabibianPRA2011}.

\subsection{Discussion} \label{lowSat}

When the cavity mode is resonant with the driving field, i.e., $\delta_c\sim0$,  the dynamics is dominated by the squeezing term in the rhs of Eq.~\eqref{Heff}. This case has been extensively discussed in Ref.~\cite{HabibianPRA2011}. Here, instead, we will consider the regime of parameters for which the dynamics is dominated by the non-degenerate parametric-amplifier type of interaction. This regime is found when the frequency of the pump is tuned between the spins and the cavity frequency, such that
\begin{eqnarray}\label{deltacEQdeltab}
\delta_c\sim -\delta_b\,,
\end{eqnarray}
and warrants that the Hamiltonian term proportional to $\alpha_{Q,a}$ in Eq.~\rp{Heff} describes the coupling between resonant transitions, as long as the four-photon processes in~\eqref{Heff} scaling with $\chi$ can be neglected. This requirement is satisfied when the number of atoms is sufficiently large. In this limit, the Hamiltonian becomes quadratic in the spin-wave and cavity mode operators $b_Q$ and $a$.

We now discuss the validity of the low-saturation assumption, on which the effective dynamics in Eq. \eqref{Master:Eq} is based.  For this purpose we estimate the atomic excited state population for atom $j$, $\av{S_j\da S_j}\sim\av{b_j\da b_j}$, and check when it is smaller than unity. This check is performed for all $j=1,\ldots,N$, where here $\av{S_j\da S_j}={\rm Tr}\{\varrho(t)S_j\da S_j\}$. In order to estimate $\av{S_j\da S_j}\sim\av{b_j\da b_j}$, we first remind that mode $b_{Q'}$ is not directly coupled to the cavity and is only non-resonantly coupled with the other spin-wave modes. Therefore, the populations of the spin-wave modes with $Q''\neq Q$ is always very small. The spin-wave mode $b_Q$, on the other hand, can be pumped by the non-linear processes in Eq.~\eqref{Heff}. We then write
\begin{eqnarray}\label{lowsat:0}
\av{b_j\da b_j}\simeq \frac{1}{N}\sum_{q}\av{b_q\da b_q}\simeq\frac{1}{N}\pq{\av{b_{Q'}\da b_{Q'}}+\av{b_Q\da b_Q}}\,,
\end{eqnarray}
where in the last passage we have neglected the populations of the spin-wave modes with $Q''\neq Q,Q'$ as well as  correlations between spin modes. Using Eq.~\eqref{b:Q} in Eq. \eqref{lowsat:0}, the latter can be reduced to the form
\begin{eqnarray}\label{lowsat}
\av{b_j\da b_j}\simeq\frac{\Omega^2}{\omega_z^2}+\frac{\av{b_Q\da b_Q}}{N}\ll 1\,,
\end{eqnarray}
which is smaller than unity in the low saturation limit here assumed and provided that $\av{b_Q\da b_Q}\ll N$, as is the case.

\section{Gaussian dynamics, two-mode squeezing, and entanglement}\label{Sec:Gauss}

We are interested in identifying the parameter regimes for which the spin-wave and the cavity mode are two-mode squeezed by the coupled dynamics described by Hamiltonian~\eqref{Heff}. The figure of merit by which we characterize the quantum correlations between the two modes is the logarithmic negativity, a measure of bipartite entanglement for Gaussian states that stems directly from the positive partial transpose (PPT) criterion of separability~\cite{Vidal,Plenio,Adesso}. In general, the dynamics described by the effective Hamiltonian~\eqref{Heff} is not Gaussian, due to the four-photon processes occurring at rate $\chi$. On the other hand, for a sufficiently large number of atoms $N$ and small number of excitations, these processes are negligible in comparison with the other terms in Eq.~\eqref{Heff} and the Hamiltonian is approximated by a quadratic form in the field operators. In this limit, master equation \eqref{ME} preserves Gaussianity, and complete information about the system is contained in the equations for the first- and second moments of the field operators~\cite{Adesso}.

In this section we analyze the limit in which the four-photon processes are negligible, and we introduce the physical quantities that we are going to evaluate, such as the logarithmic negativity and the spectrum of two-mode squeezing. We also analyze the conditions under which the dynamics is equivalent to the one of a parametric amplifier. These analytical studies provide a basis for identifying the regimes of parameters for which two-mode squeezing and bipartite entanglement are expected to arise. These are the regimes that we are going to consider when integrating master equation \eqref{Master:Eq}.

\subsection{Gaussian dynamics and covariance matrix}

We first implement a compact representation of the Heisenberg-Langevin equations by introducing the vectors of operators
\begin{eqnarray*}
&&{\bf a}=\pt{a,b_Q,a\da,b_Q\da}^T\,,\\
&&{\bf a}^{\rm in}(t)=\pt{a_{\rm in}(t),b_{\rm in}(t),a_{\rm in}\da(t),b_{\rm in}\da(t)}^T\,,\
\end{eqnarray*}
such that the Heisenberg-Langevin equations \eqref{HL:a}-\eqref{HL:b} can be cast in the form
$$\dot\va(t)=\ZZ\,\va(t)+\QQ\,\va^{\rm in}(t)\,.$$
Here, ${\mathcal Z}={\mathcal Z}_0+{\mathcal Z}_1$, with
\begin{eqnarray}\label{Z}
&&{\mathcal Z}_0=-\ii\left(\begin{array}{llll}
\delta_c & \alpha_{\rm BS} & 0 & \alpha_{Q,a} \\
\alpha_{\rm BS} & \delta_b& \alpha_{Q,a} & 2\alpha_{Q} \\
0 & -\alpha_{Q,a} & -\delta_c & -\alpha_{\rm BS} \\
-\alpha_{Q,a} & -2\alpha_{Q} & -\alpha_{\rm BS} & -\delta_b \\
 \end{array}\right)\,,\\
&&{\mathcal Z}_1=-\left(\begin{array}{llll}
\kappa& 0 & 0 & 0 \\
0 & \gamma/2& 0 & 0 \\
0 & 0 & \kappa & 0 \\
0 & 0 & 0 & \gamma/2 \\
 \end{array}\right)\,,
\end{eqnarray}
and $\QQ$ is a diagonal matrix with elements $(\sqrt{2\kappa},\sqrt{\gamma},\sqrt{2\kappa},\sqrt{\gamma})$. For later convenience we also introduce the quadrature operators $x_a=(a+a\da)$, $p_a={\rm i}(a\da-a)$, and the corresponding quadratures $x_b$ and $p_b$ constructed from operators $b_Q$ and $b_Q\da$. The vector of quadrature operators takes the form $${\bf x}=\pt{x_a,p_a,x_b,b_b}^T=\Pi\ {\bf a}\,,$$ with
\begin{equation}
\Pi=\left(\begin{array}{llll}
1 & 0 & 1 & 0 \\
-\ii & 0 & \ii & 0 \\
0 & 1 & 0 & 1 \\
0 & -\ii & 0 & \ii \\
\end{array}\right)\,.
\end{equation}
We further introduce the covariance matrix $\mathcal C$, defined as
\begin{equation}
\label{C}
{\cal C}=\pq{\av{{\bf x}{\bf x}^T}+\av{{\bf x}{\bf x}^T}^T}/{2}-\av{{\bf x}}\av{{\bf x}}^T\,.
\end{equation}

Using Eqs. \eqref{HL:a} and \eqref{HL:b} we find the equations for the first moment $ \av{{\bf a}}$ as well as for the second moments,
$${\cal A}_{j,k}=\av{{\bf a}_j{\bf a}_k}\,,$$
corresponding to the elements of the correlation matrix ${\cal A}=\av{{\bf a}\,{\bf a}^T}$. Their equations of motion read
\begin{eqnarray}\label{gaussianEv}
\av{\dot {\bf a}}&=&{\cal Z} \av{{\bf a}}\,,
\nn\\
\dot{\cal A}&=&{\cal Z}{\cal A}+{\cal A}{\cal Z}^T+{\cal N}\,,
\end{eqnarray}
where
\begin{equation}
{\cal N}=\left(\begin{array}{llll}
0 & 0 & 2\kappa & 0 \\
0 & 0 & 0 & \gamma \\
0 & 0 & 0 & 0 \\
0 & 0 & 0 & 0 \\
 \end{array}\right)\,.
\end{equation}
Solving the equations of motion for $ \av{{\bf a}}$ and ${\cal A}$ gives the full characterization of the system dynamics, provided the initial state is Gaussian. In particular, from $ \av{{\bf a}}$ and ${\cal A}$ one finds the covariance matrix at all instants of time, since using Eq. \eqref{C} it can be cast in the form
\begin{eqnarray}\label{C:1}
{\cal C}=\Pi \frac{{\cal A} +{\cal A}^{\rm T}}{2}\Pi^{\rm T} -\Pi \av{\bf a}\av{\bf a}^T\Pi^{\rm T}\,. \end{eqnarray}

\subsection{Non-degenerate parametric amplification}

Our purpose is to identify the regime of parameters of the coherent dynamics where the latter can be characterized as non-degenerate parametric amplification. We thus analyze the matrix ${\cal Z}_0$, which determines the coherent dynamics in the Gaussian limit. Its eigenvalues read
\begin{eqnarray}\label{lambda}
\lambda_{1\pm}&=&\mp\ii\sqrt{\Lambda_1+\sqrt{\Lambda_2}}\,,\nn\\
\lambda_{2\pm}&=&\mp\ii\sqrt{\Lambda_1-\sqrt{\Lambda_2}}\,,
\end{eqnarray}
with
\begin{eqnarray}
\Lambda_1&=&{\frac{\delta_c^2+\delta_b^2}{2} +\alpha_{BS}^2-\alpha_{Q,a}^2-2\alpha_Q^2}\,,\\
\Lambda_2&=&4\lpq{\pt{\alpha_Q^2+\frac{\delta_c^2-\delta_b^2}{4}}^2 -
2\alpha_{BS}\alpha_{Q,a}\alpha_Q\delta_c
}\label{Lambda:2}\\
&&\rpq{
+\alpha_{Q,a}^2\pt{\alpha_Q^2-\frac{\pt{\delta_c-\delta_b}^2}{4}}+\alpha_{BS}^2\pt{\frac{\pt{\delta_c+\delta_b}^2}{4}-\alpha_Q^2}
 }.\nn
\end{eqnarray}
Here, $\Lambda_1>0$. In fact, the model we have derived is based on the assumption that the frequencies $\delta_c$ and $\delta_b$ are the largest parameters (being $|\delta_c|,|\delta_b|\sim |\omega_z|/2$), and such that $|\delta_c-\delta_b|\sim |\omega_z|$, as one can see from Eq.~\eqref{deltacEQdeltab}. Under the assumption $|\delta_c+\delta_b|\ll|\delta_c-\delta_b|$, then $\Lambda_1\gg\sqrt{|\Lambda_2|}$. The frequencies characterizing the eigenvalues are the frequencies of the normal modes and are given in leading order by
$$\Delta_{\pm} \approx \pm \sqrt{\Lambda_1}\sim \pm (\delta_b-\delta_c)/2\,,$$ so that the splitting between the normal modes is of the order of $ |\delta_b-\delta_c|$.

With these results at hand it is possible to distinguish two regimes: One is found when the eigenvalues are purely imaginary, namely, for $\Lambda_2>0$. In this case the dynamics is periodic in time. When instead  $\Lambda_2<0$, all four eigenvalues have a real part and the total number of excitations can increase with time. In detail, after expanding in first order in the parameter $1/|\delta_b-\delta_c|$ (after assuming $|\delta_b-\delta_c|\gg|\alpha_Q|,|\alpha_{Q,a}|,|\alpha_{BS}|,|\delta_b+\delta_c|$) we find
\begin{eqnarray}
\lambda_{1\pm}&=&\mp\ii\pq{\frac{\delta_b-\delta_c}{2}+\frac{1}{2}\sqrt{
(\delta_b+\delta_c)^2-4\alpha_{Q,a}^2}}+{\cal O}\pt{1/|\delta_b|}\,,
\nn\\
\lambda_{2\pm}&=&\mp\ii\pq{\frac{\delta_b-\delta_c}{2}-\frac{1}{2}\sqrt{
(\delta_b+\delta_c)^2-4\alpha_{Q,a}^2}}+{\cal O}\pt{1/|\delta_b|}\,.\nn\\
\end{eqnarray}
Parametric amplification is here found for $|2\alpha_{Qa}|>|\delta_b+\delta_c|$.

Let now $\Lambda_2<0$,  at leading order in $1/\Lambda_1$, the real part of the eigenvalues in Eq.~\rp{lambda} is $\pm\frac{1}{2}\sqrt{\abs{\Lambda_2}/\Lambda_1}$. If one now also considers the dissipative processes and takes for convenience $\kappa=\gamma/2$, the result is modified by adding a real term equal to $-\kappa$ to each eigenvalue. 
In analogy with the physics of the parametric oscillator~\cite{Reid1990,Reid2004,MilburnWalls} the dissipative dynamics of our gaussian model is characterized by a threshold value for the strength of the non-linearities above which the populations of the modes diverge. Specifically 
parametric amplification is 
below or above threshold depending on whether 
$ \frac{1}{2}\sqrt{\abs{\Lambda_2}/\Lambda_1}$ 
is smaller or larger than $\kappa$. The expressions become more involved when $\gamma\neq 2\kappa$, and their analytical study does not provide further insight for our purposes so that we refer the interested reader to the work in Refs.~\cite{Reid1990,Reid2004}. 
 We remark that above and close to the threshold, the validity of our gaussian model is questionable. Indeed, when the population of the spin wave mode is too large the four-photon process are not negligible and, 
more in general, the low saturation expansion can fail as discussed in Sec.~\ref{lowSat}.
In the following, when including dissipation, we will focus on the regime in which the system operates below threshold, so that a stationary state of the dynamics can be determined.

\subsection{Two-mode squeezing}

Non-degenerate parametric amplification is known to generate two-mode squeezing. The Hamiltonian in Eq.~\eqref{Heff}, however, contains further terms which can compete against the dynamics dictated by the parametric-amplifier term even after setting $\chi=0$. In this paper we analyze the conditions and parameters for which two-mode squeezing between the cavity and the spin-wave modes is found.

The system is said to be two-mode squeezed when the minimum of the variance $\Delta X_t (\theta_a,\theta_b)$ of the composite quadrature
\begin{equation}
\label{X:t}
X_t(\theta_a,\theta_b)=(a(t)\ee^{\ii\theta_a}+a\da(t)\ee^{-\ii\theta_a}
+b(t)\ee^{\ii\theta_b}+b\da(t)\ee^{-\ii\theta_b})/\sqrt{2}
\end{equation}
 is below the shot-noise limit for some angles $\theta_a$ and $\theta_b$. Thus, given
\begin{eqnarray}\label{DeltaX}
\Delta X(\theta_a,\theta_b)=\av{X(\theta_a,\theta_b)^2}-\av{X(\theta_a,\theta_b)}^2\,,
\end{eqnarray}
the condition for two-mode squeezing can be summarized by the inequality
\begin{equation}
\label{Two:mode}
{\rm min}_{\theta_a,\theta_b}\{\langle\Delta X_t (\theta_a,\theta_b)\rangle\}<1\,,
\end{equation}
where the expectation values are taken over the density matrix of the composite system at time $t=0$ and the shot-noise limit is unity from the definition in Eq.~\eqref{X:t}. This condition can be fulfilled at some instants of time and/or at asymptotic times of the dynamics. In this latter case two-mode squeezing is a property of the stationary state and can be experimentally revealed by analyzing the spectrum of squeezing of the fields emitted by the atoms and by the cavity field.

Let us assume that the field spontaneously emitted by the atoms can be efficiently collected, such as by means of an additional cavity~\cite{ParkinsKimble}.
This would allow one to perform balanced heterodyne detection using two local oscillators at frequencies $\omega_{LOa}$ and $\omega_{LOb}$, interfering with the cavity and the atoms output fields, respectively, at a beam splitter~\cite{Reid1989}. We denote the output field containing the information on the spin mode by $b_{\rm out}^{(d)}$, where the superscript $(d)$ indicates that the detected output fields can be a fraction of the total emitted field at time $t$ due to the finite detector collection efficiency. We also denote by $a_{\rm out}^{(d)}$ the field at the cavity output which can be collected at the corresponding detector and consider the composite quadrature of the output fields which are ideally measured at the detector,
\begin{eqnarray}
X_{\rm out}\al{\theta_a,\theta_b,\xi}(t)&=&\frac{\abs{\xi}}{\sqrt{\xi^4+1}}\lpq{\abs{\xi}a_{\rm out}^{(d)}(t)\ee^{\ii\pt{\theta_a+\Delta_a t}}+\abs{\xi}a_{\rm out}^{\dagger(d)}(t)\ee^{-\ii\pt{\theta_a+\Delta_a t}}
}\nn\\&&\rpq{
+\frac{1}{\xi}b_{\rm out}^{(d)}(t)\ee^{\ii\pt{\theta_b+\Delta_b t}}+\frac{1}{\xi}b_{\rm out}^{\dagger(d)}(t)\ee^{-\ii\pt{\theta_b+\Delta_b t}}
}\nn\,,\\
\label{X:out}
\end{eqnarray}
where $\Delta_a=\omega_{LOa}-\omega_p$ and $\Delta_b=\omega_{LOb}-\omega_p$ indicate the frequencies of the local oscillators relative to the driving laser frequency, while $\theta_a$ and $\theta_b$ are the local oscillators phases which can be independently adjusted. The Fourier transform of the fluctuations of the difference photocurrent induced at the detectors (here weighted by the real parameter $\xi$) gives the spectrum of squeezing, which is defined as~\cite{MilburnWalls}
\begin{eqnarray}
\label{Squeeze}
 S_\xi\al{\theta_a,\theta_b}(\omega)=2{\rm Re}\int_0^\infty\ \dd t\ \ee^{-\ii\omega t}
\av{X_{\rm out}\al{\theta_a,\theta_b,\xi}(t)\ X_{\rm out}\al{\theta_a,\theta_b,\xi}(0)}_{st}\,.
\end{eqnarray}
Here, the expectation values are taken over the density matrix of the system at steady state and we have used that $\av{X_{\rm out}\al{\theta_a,\theta_b,\xi}(t)}_{st}=0$ at any time $t$. A detailed calculation of the spectrum of squeezing is reported in the Appendix.

In the form given in Eq.~\eqref{Squeeze}, the squeezing spectrum depends on the parameter $\xi$. This parameter has to be varied in order to check whether for some $\xi$ the inequality is satisfied:
\begin{equation}
\label{S:sum}
S_\xi\al{\theta_a,\theta_b}(\omega)+S_\xi\al{\theta_a+\frac{\pi}{2},\theta_b-\frac{\pi}{2}}(\omega)<2\,,
\end{equation}
which is a necessary and sufficient condition for entanglement of Gaussian states~\cite{Vitali2006}.

We now comment on the finite collection efficiency at the detectors. Here, it is introduced by means of the detected output fields $a_{\rm out}^{(d)}(t)$ and $b^{(d)}_{\rm out}(t)$, and their adjoints, which are related to the input-noise operators through the relations
\begin{eqnarray}\label{aout2}
 && a_{\rm out}^{(d)}(t)=\sqrt{2\kappa}\,a(t)-a_{\rm in}^{(d)}(t)\,,\nn\\
 && b_{\rm out}^{(d)}(t)=\sqrt{\gamma}\,b_Q(t)-b^{(d)}_{\rm in}(t)\,,
\end{eqnarray}
where $a^{(d)}_{\rm in}(t)$ and $b^{(d)}_{\rm in}(t)$ only include the modes within the collection efficiency of the detector. We denote by $a^{(nd)}_{\rm in}(t)$ and $b^{(nd)}_{\rm in}(t)$ the input-noise operators for the modes which are not detected. Using this partition, the Heisenberg-Langevin equations in Eqs.~\eqref{HL:a}-\eqref{HL:b} can be cast in the form
\begin{eqnarray}
\label{HL:a:nd}
&&\dot{a}=\frac{\ii}{\hbar}[{\mathcal H}_{\rm eff},a]-\kappa a+ \sqrt{2\kappa'}a_{\rm in}^{(d)}+\sqrt{2\kappa''}a_{\rm in}^{(nd)}\,,\\
&&\dot{b}_Q=\frac{\ii}{\hbar}[{\mathcal H}_{\rm eff},b_Q]-\frac{\gamma}{2} b_Q+\sqrt{\gamma'}b_{\rm in}^{(d)}+\sqrt{\gamma''}b_{\rm in}^{(nd)}\,,
\label{HL:b:d}
\end{eqnarray}
where $\kappa'=\eta_a\kappa$ and $\gamma'=\eta_b\gamma$, with $0\le \eta_a,\eta_b\le 1$ the collection efficiencies, while $\kappa''=\kappa(1-\eta_a)$ and $\gamma''=\gamma(1-\eta_b)$. Since detected and non-detected modes are here orthogonal (they possess, for instance, different directions of propagation), the corresponding input and output field operators are uncorrelated one from the other. In detail, $\av{a_{\rm in}^{(\ell)}(t)}=\av{b_{\rm in}^{(\ell)}(t)}=0$ and $\av{a_{\rm in}^{(\ell)}(t)a^{\dagger(\ell')}_{\rm in}(t')}=\av{b_{\rm in}^{(\ell)}(t)b^{\dagger(\ell')}_{\rm in}(t')}=\delta_{\ell,\ell'}\delta(t-t')$, with $\ell,\ell'=d,nd$.

\begin{figure*}[t!]
\centering
\includegraphics[width=5.5cm]{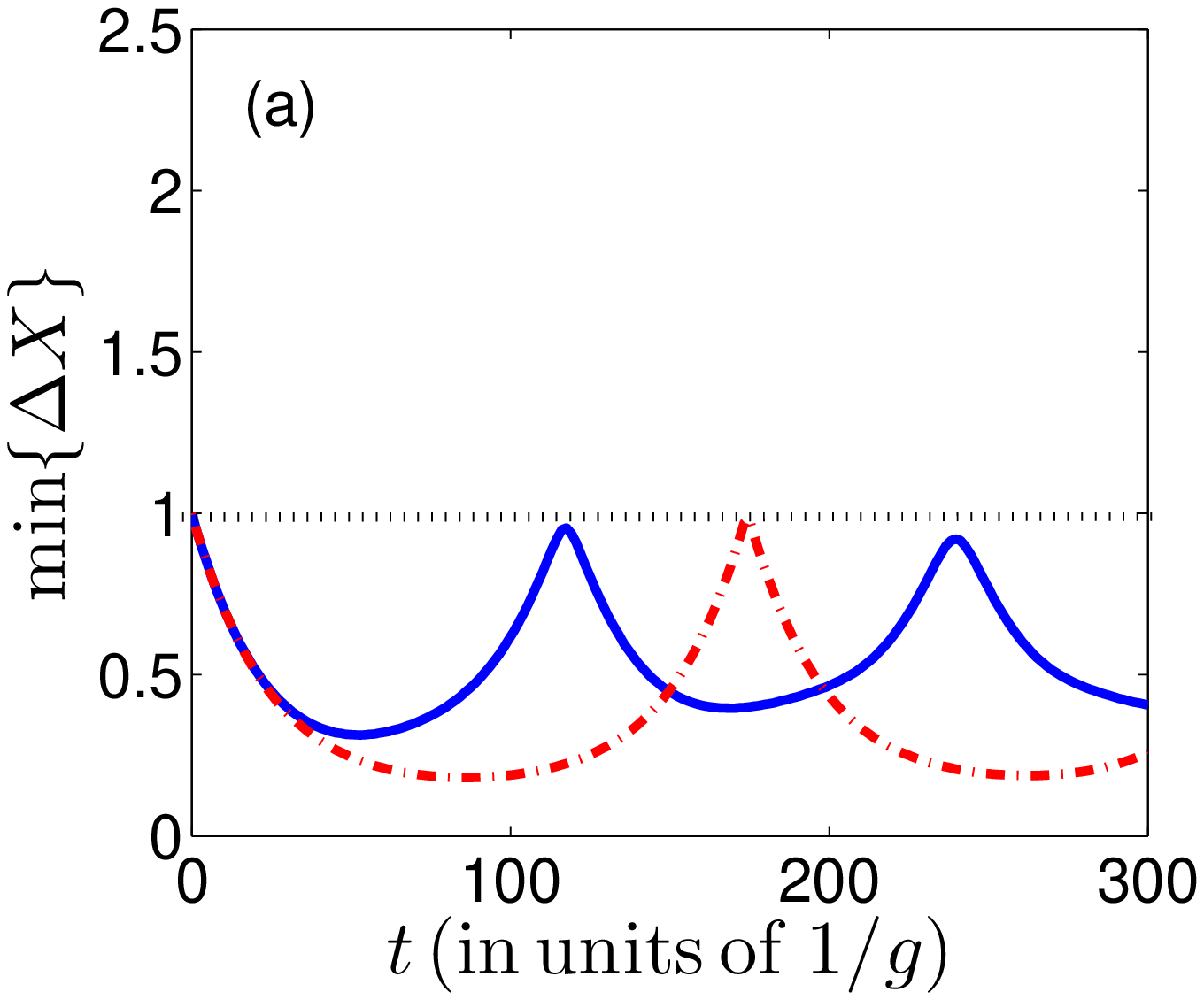}
\includegraphics[width=5.5cm]{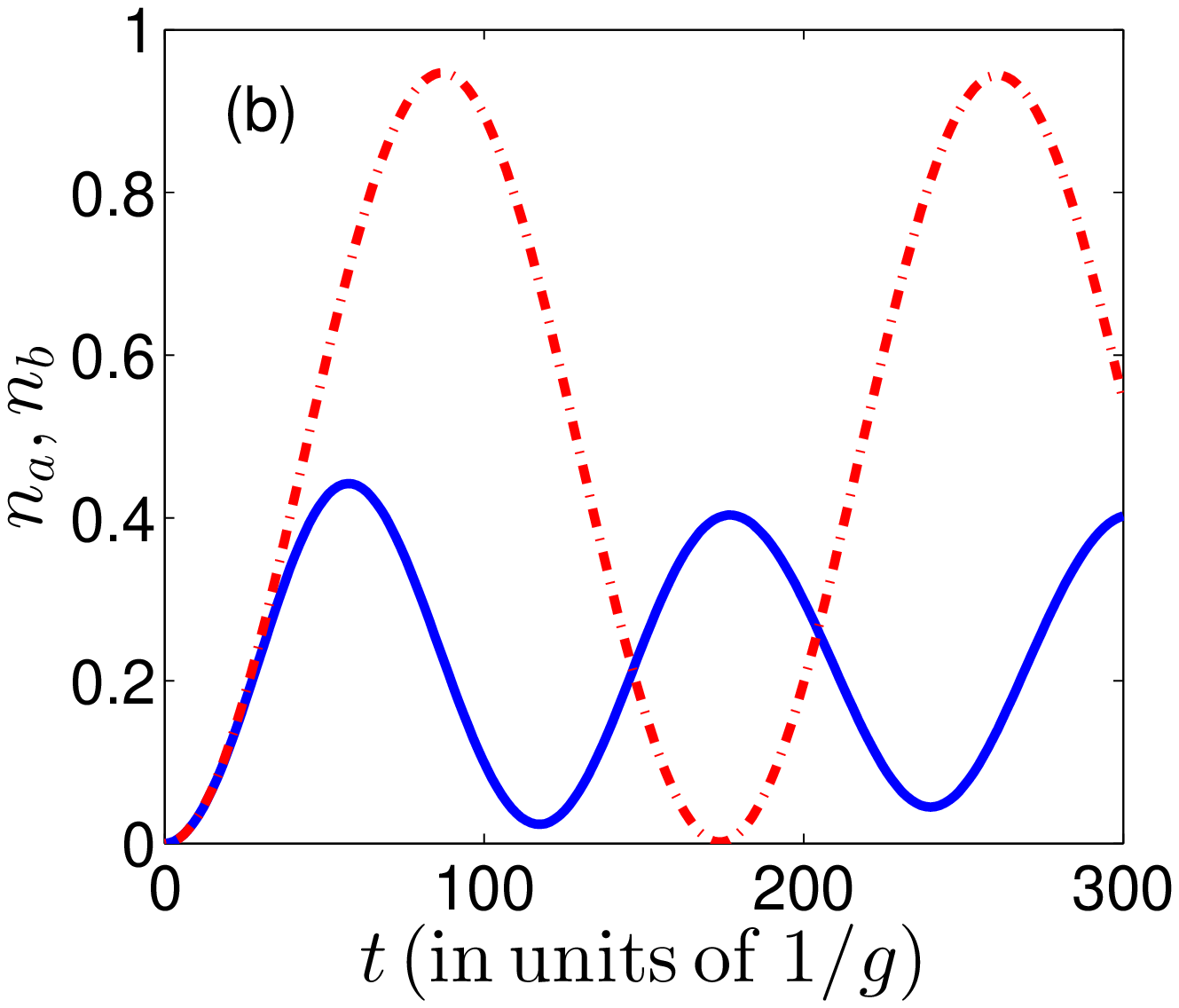}
\includegraphics[width=5.5cm]{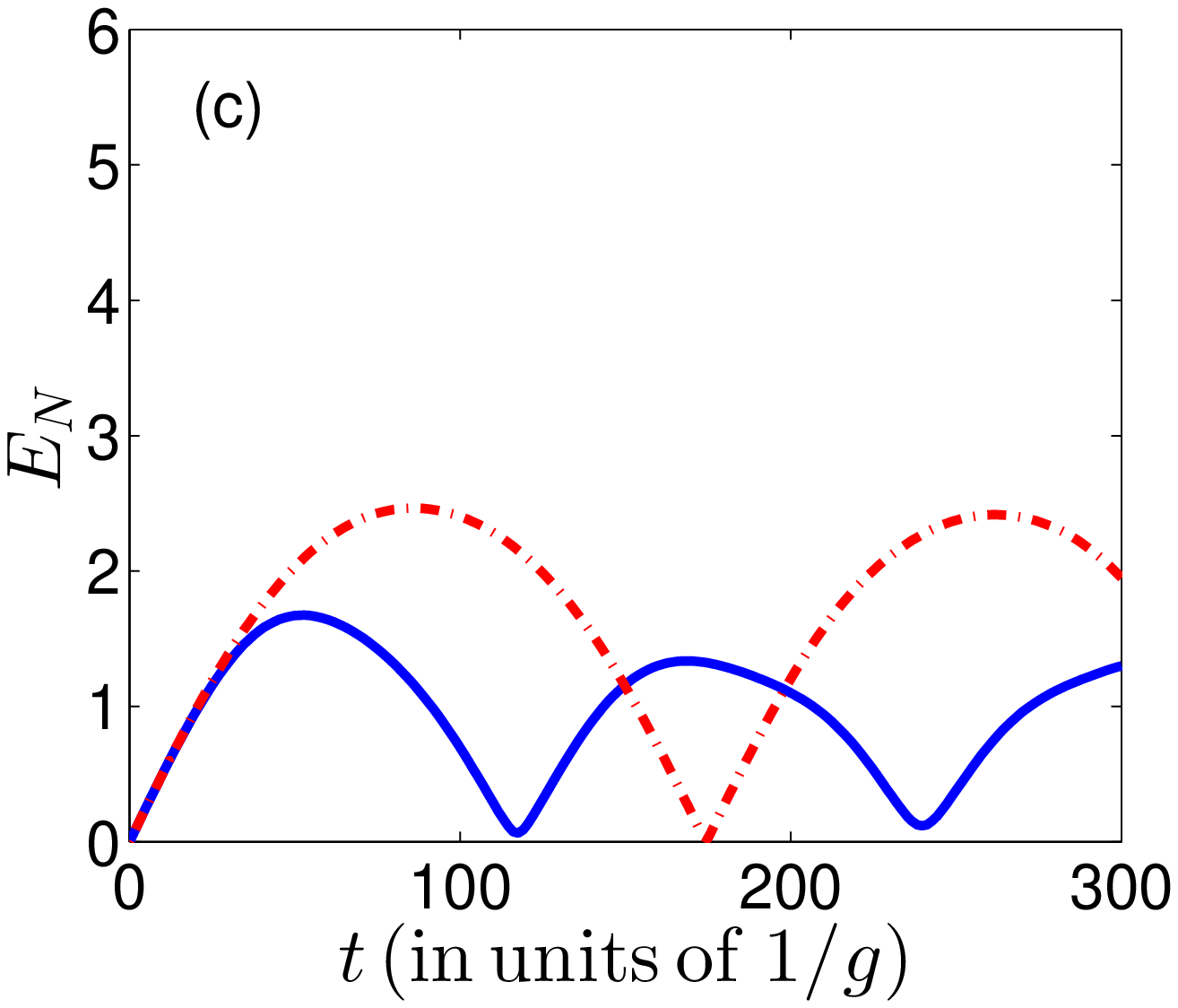}
\includegraphics[width=5.5cm]{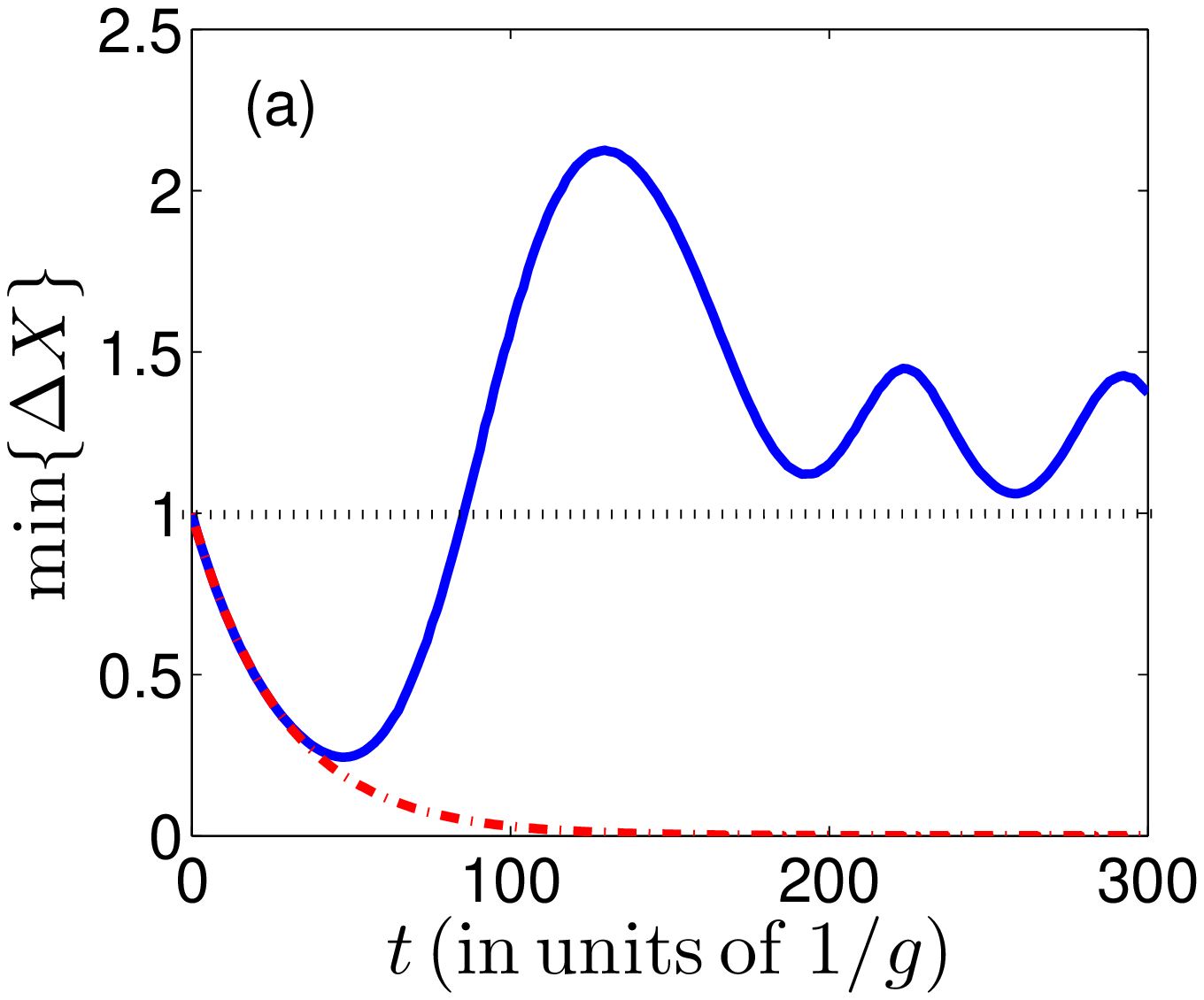}
\includegraphics[width=5.5cm]{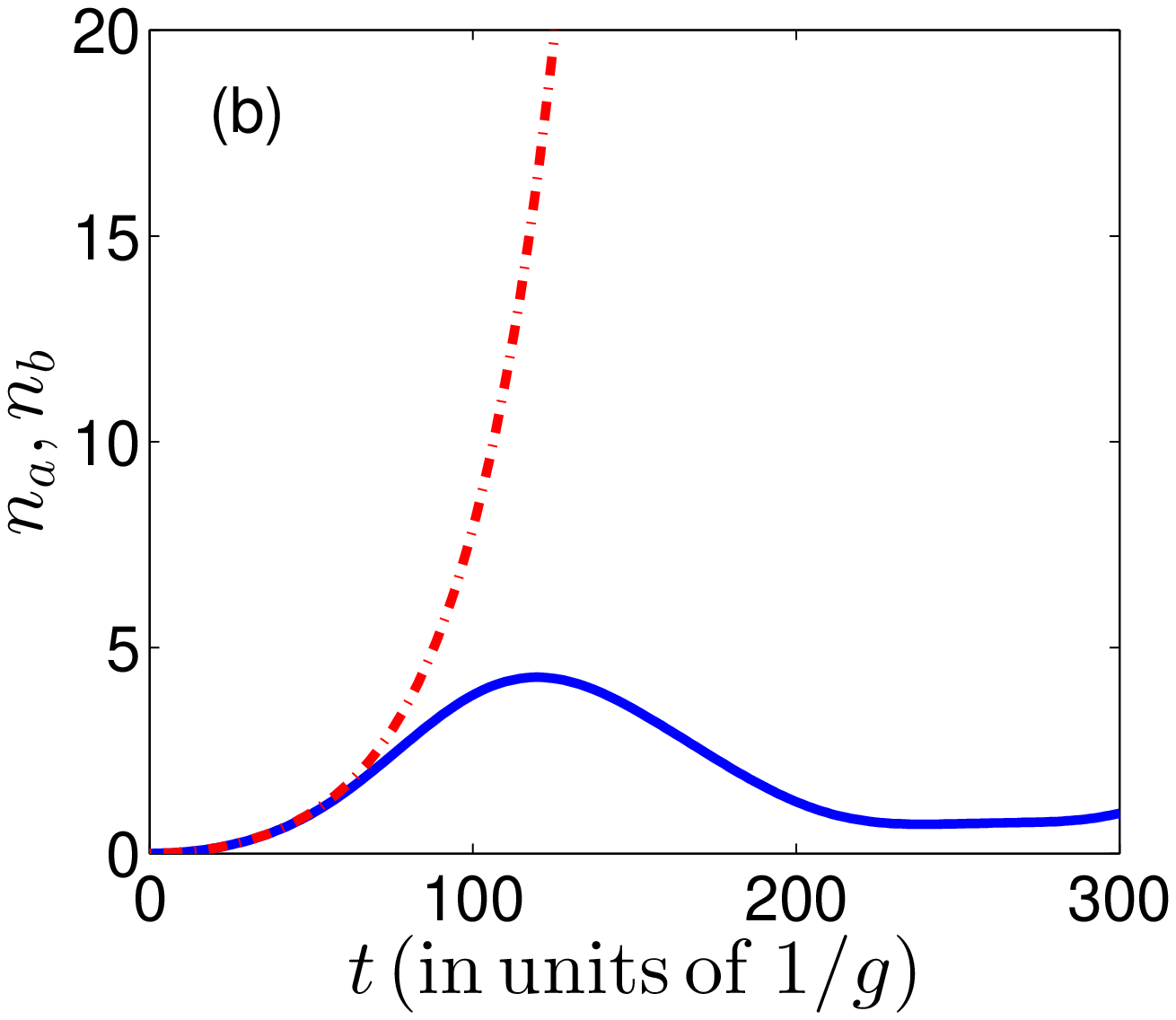}
\includegraphics[width=5.5cm]{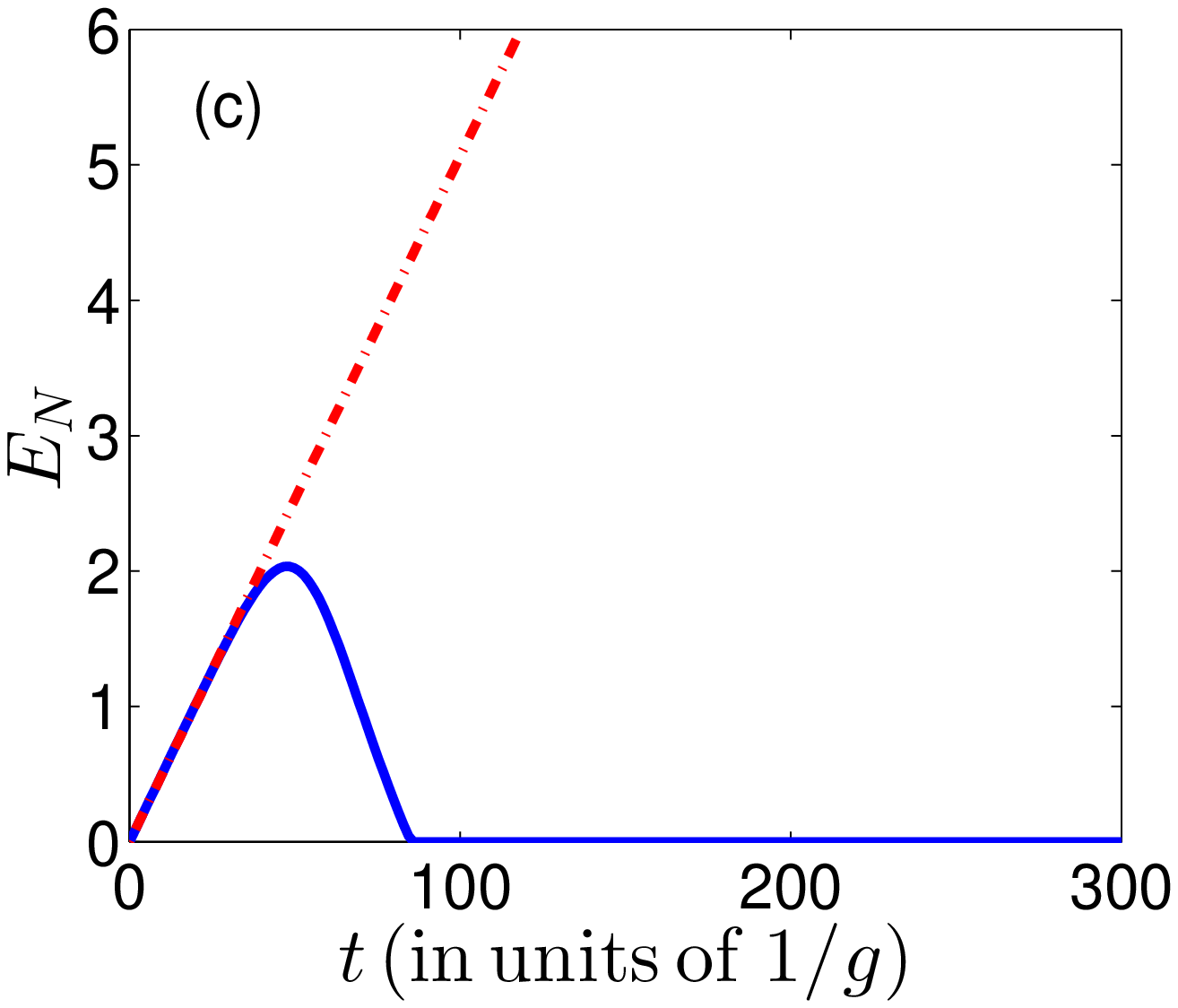}
\caption[]{(Color online) Coherent time evolution of (a) minimum of the variance of the composite quadrature, Eq.~\rp{Two:mode}, (b) population of the field modes $n_a$ and $n_b$ 
(the results for the populations are very close and the two corresponding curves are not distinguishable) 
and (c)  logarithmic negativity. The solid lines  correspond to the numerical integration of the density matrix according to the master equation, Eq.~\eqref{ME},  after setting $\kappa=\gamma=0$ and using Hamiltonian in Eq.~\rp{Heff}, the dash-dotted lines are the corresponding results setting to zero the cross-Kerr nonlinearity, namely, $\chi=0$ in Eq.~\rp{Heff}. The parameters are $N=50$ atoms and $\omega_z=100g$, $\Omega=5g$, $\delta_b=100.5g$, while $\delta_c=-100.55g$  (upper row)  and  $\delta_c=-\delta_b=-100.5g$ 
(lower row). The initial state is the vacuum for both modes. The time is in units of $1/g$, where $g$ is the vacuum Rabi frequency.}
\label{fig:2}
\end{figure*}

\subsection{Logarithmic negativity}
\label{LogNeg}

Two-mode squeezing and entanglement are two related features \cite{Reid1989,Reid2004}. In fact, for symmetric states the variance of the quadrature $\Delta X$ can be used to construct entanglement measures \cite{Adesso}. In our case, the two modes are not symmetric and the corresponding composite quadratures can be set in direct relation with entanglement after an appropriate redefinition, in which the quadrature of each field is scaled by different weights, here introduced by means of the parameter $\xi$ in the composite quadrature \eqref{X:out}. In the next section we also resort to the logarithmic negativity in order to quantify the amount of entanglement shared by atoms and light.

The logarithmic negativity $E_N$ is the logarithm of the trace norm of the partial transpose $\rho^{T_A}$ of a density matrix $\rho$ of a bipartite system, where transposition is performed with respect to one subsystem~\cite{PeresHorodeckis,Werner}. It reads $E_N=\log_2\abs{\abs{\rho^{T_A}}}$ and for a Gaussian states it can be expressed in terms of the elements of the covariance matrix by~\cite{Adesso}
\begin{equation}\label{EN}
 E_N=\max\{ 0, -\log_{2}({\nu_{-}}) \}\,,
\end{equation}
where $\nu_-$ is the smallest symplectic eigenvalue of the partially transposed covariance matrix  $$\tilde{\mathcal C}={\cal T}{\mathcal C}{\cal T}\,,$$ with
\begin{equation}
 {\cal T}=\left(\begin{array}{llll}
1 & 0 & 0 & 0 \\
0 & -1 & 0 & 0 \\
0 & 0 & 1 & 0 \\
0 & 0 & 0 & 1 \\
 \end{array}\right)\,
\end{equation}
performing the partial transposition, which is here realized by the mirror reflection $p_a\to-p_a$~\cite{Simon}. Cavity and spin-wave modes are entangled when $E_N$ is non-zero, namely when $\nu_{-}<1$. The actual value of $\nu_{-}<1$ quantifies the bipartite entanglement of Gaussian states.
In the next section we will report curves corresponding to the values of $E_N$ obtained from Eq.~\eqref{EN} by the symplectic diagonalization of $\tilde{\mathcal C}$, after evaluating the corresponding covariance matrix at a given time.

\begin{figure*}[t!]
\centering
\includegraphics[width=5.5cm]{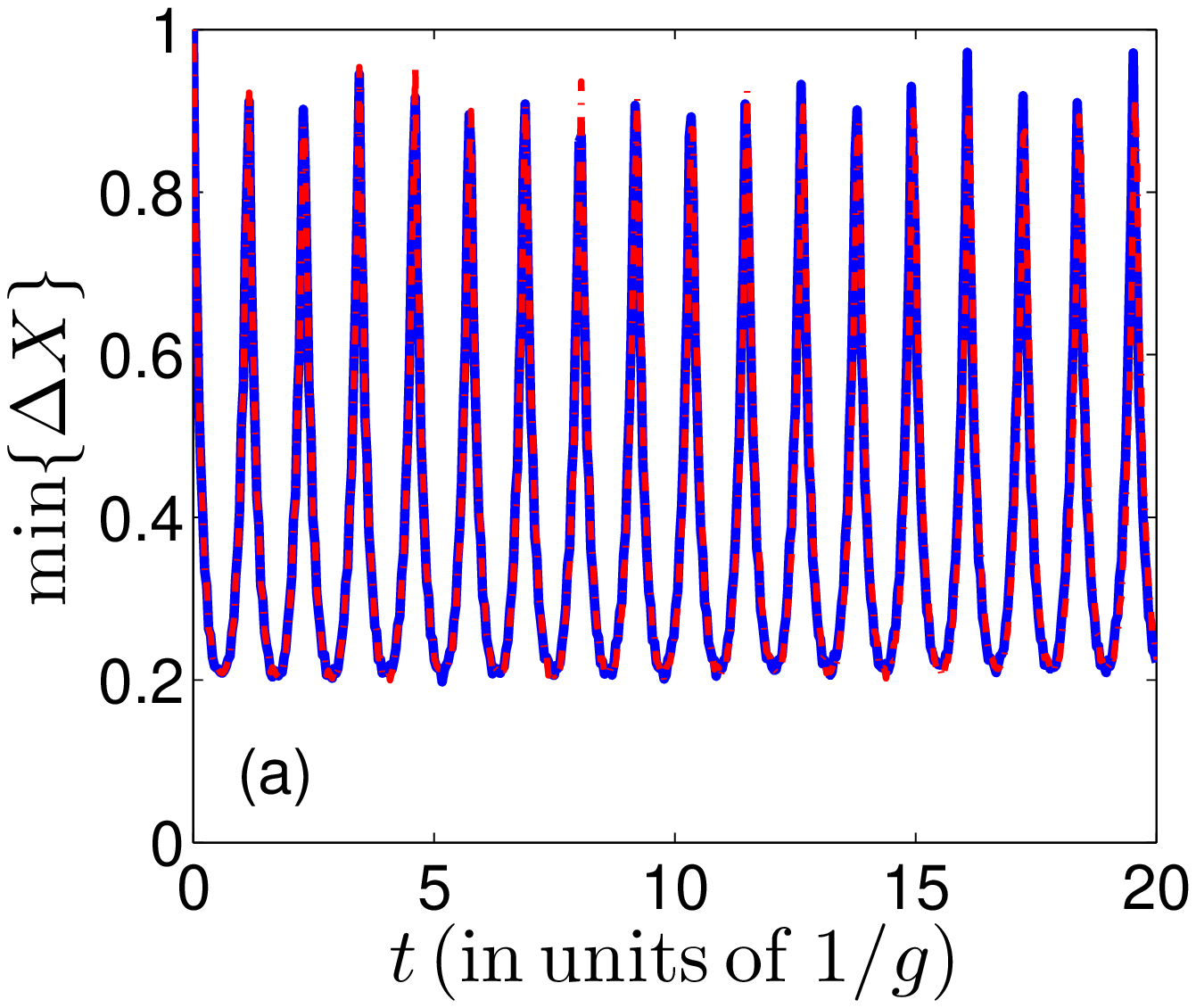}
\includegraphics[width=5.5cm]{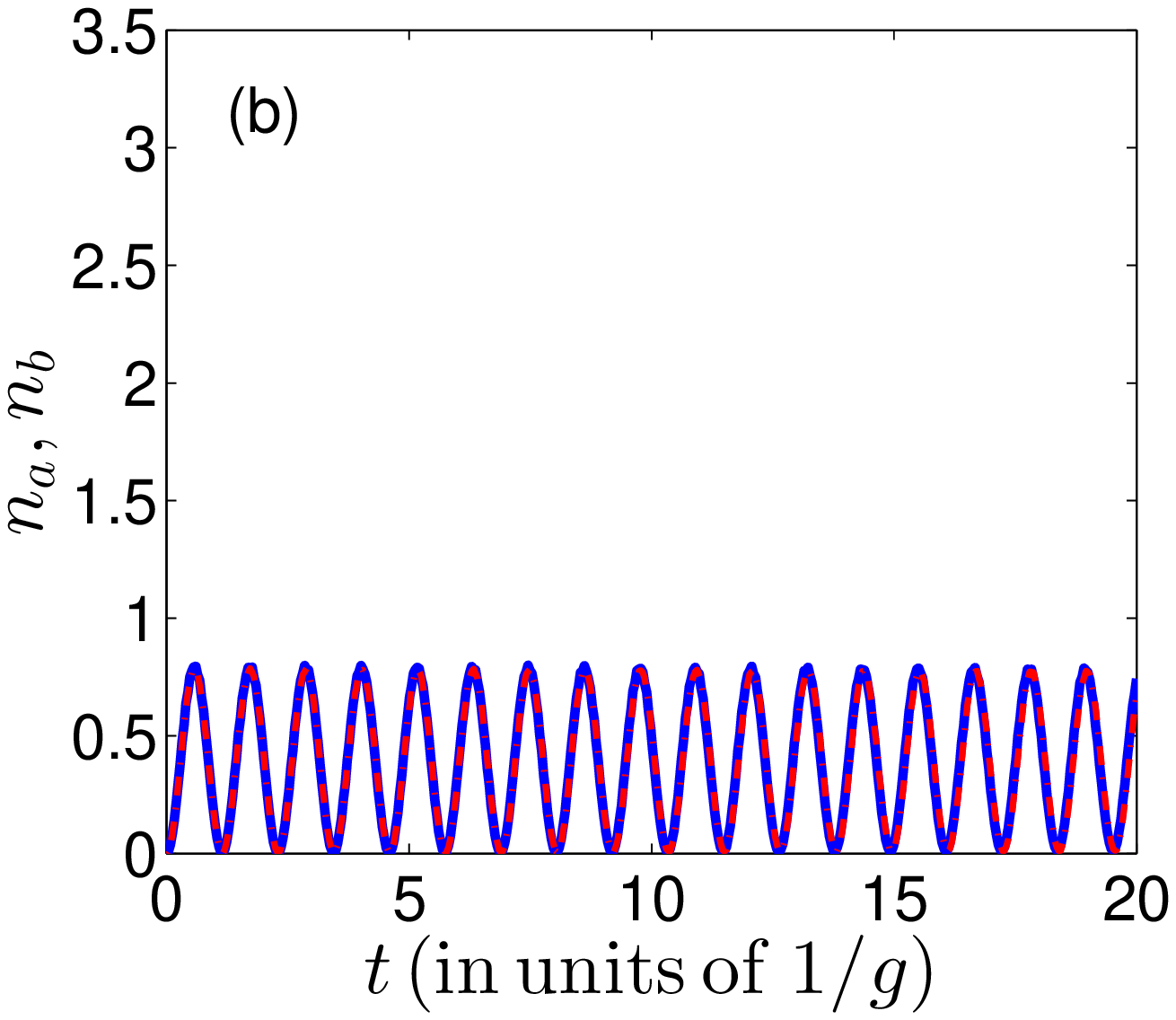}
\includegraphics[width=5.5cm]{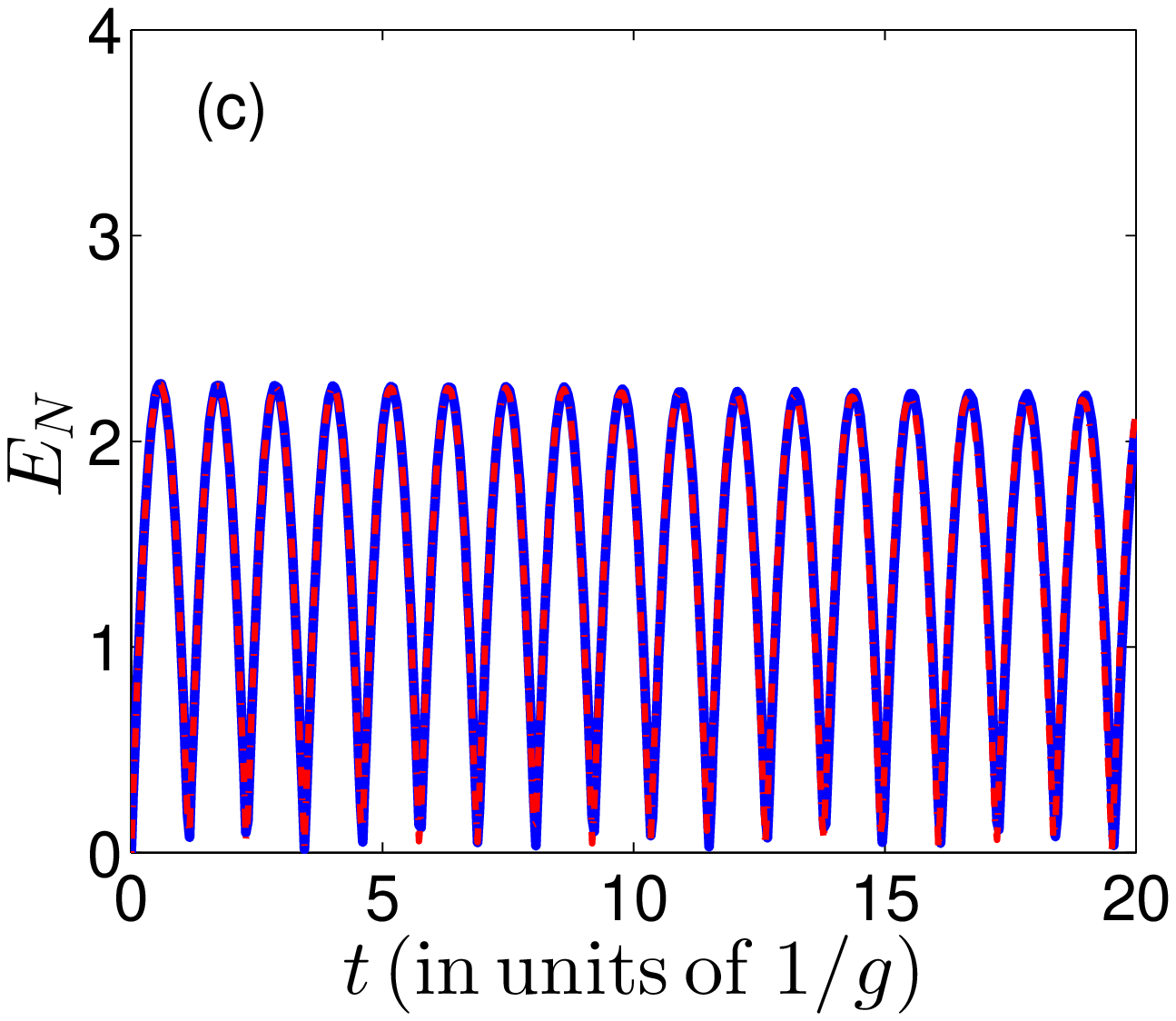}
\includegraphics[width=5.5cm]{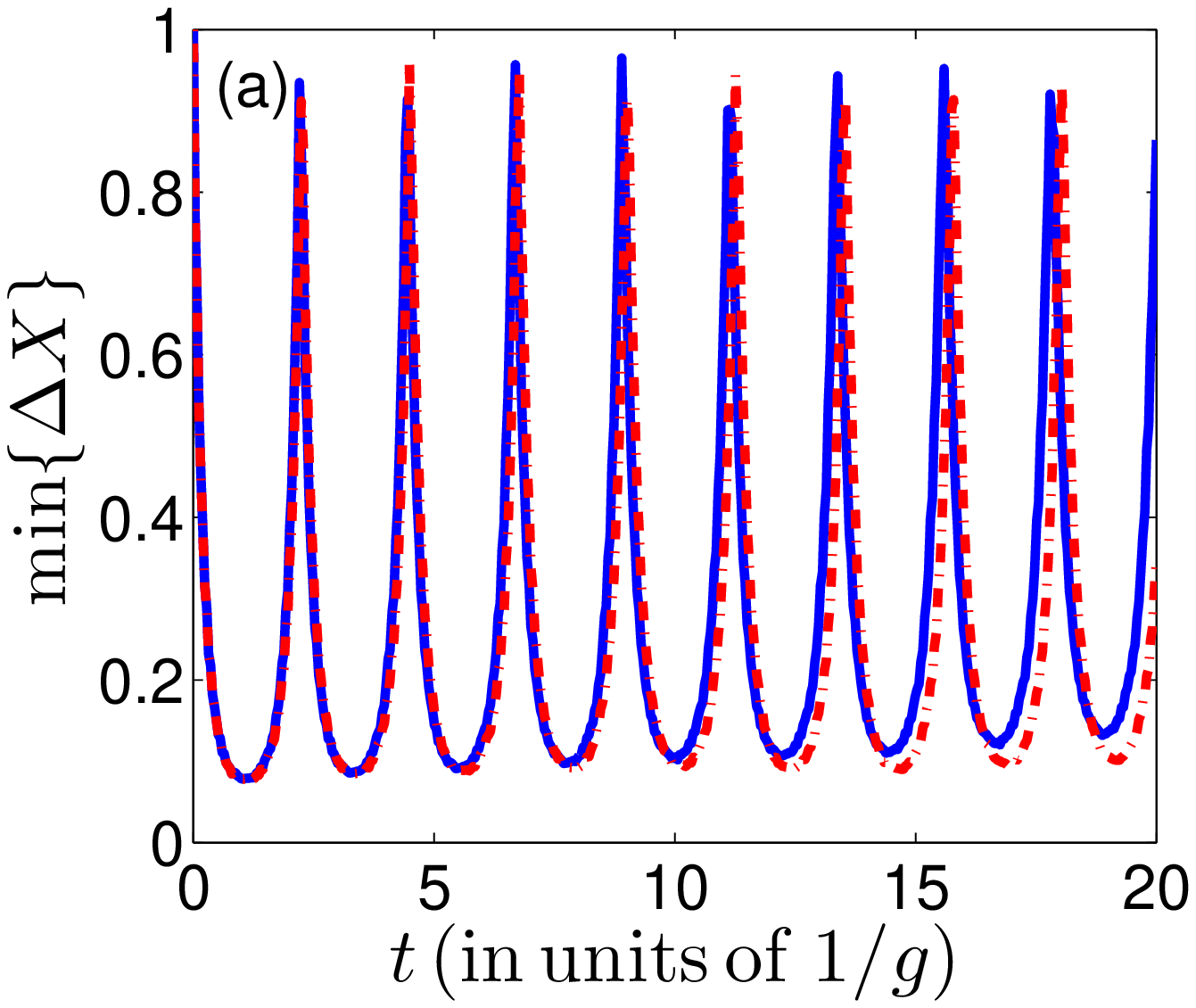}
\includegraphics[width=5.5cm]{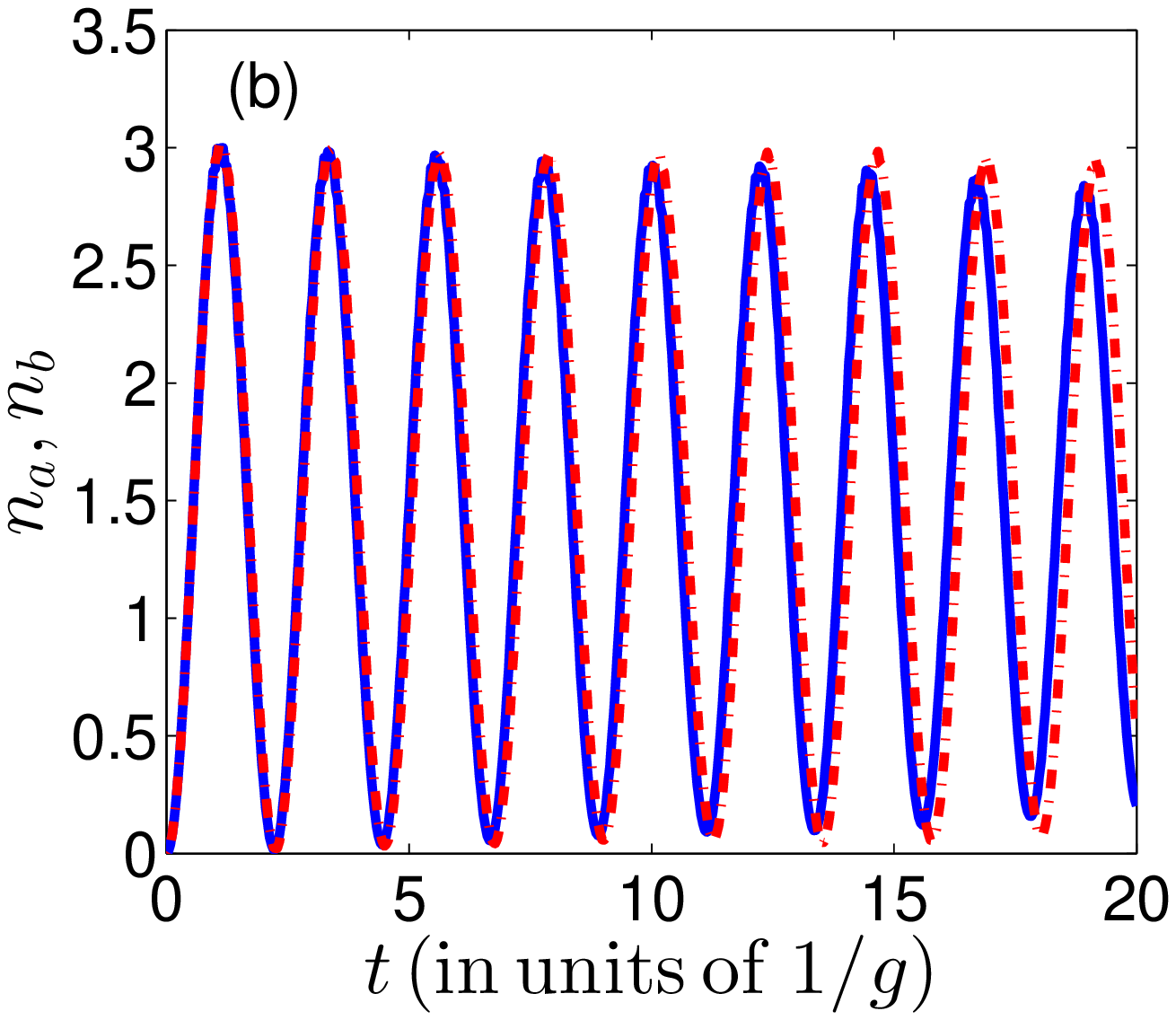}
\includegraphics[width=5.5cm]{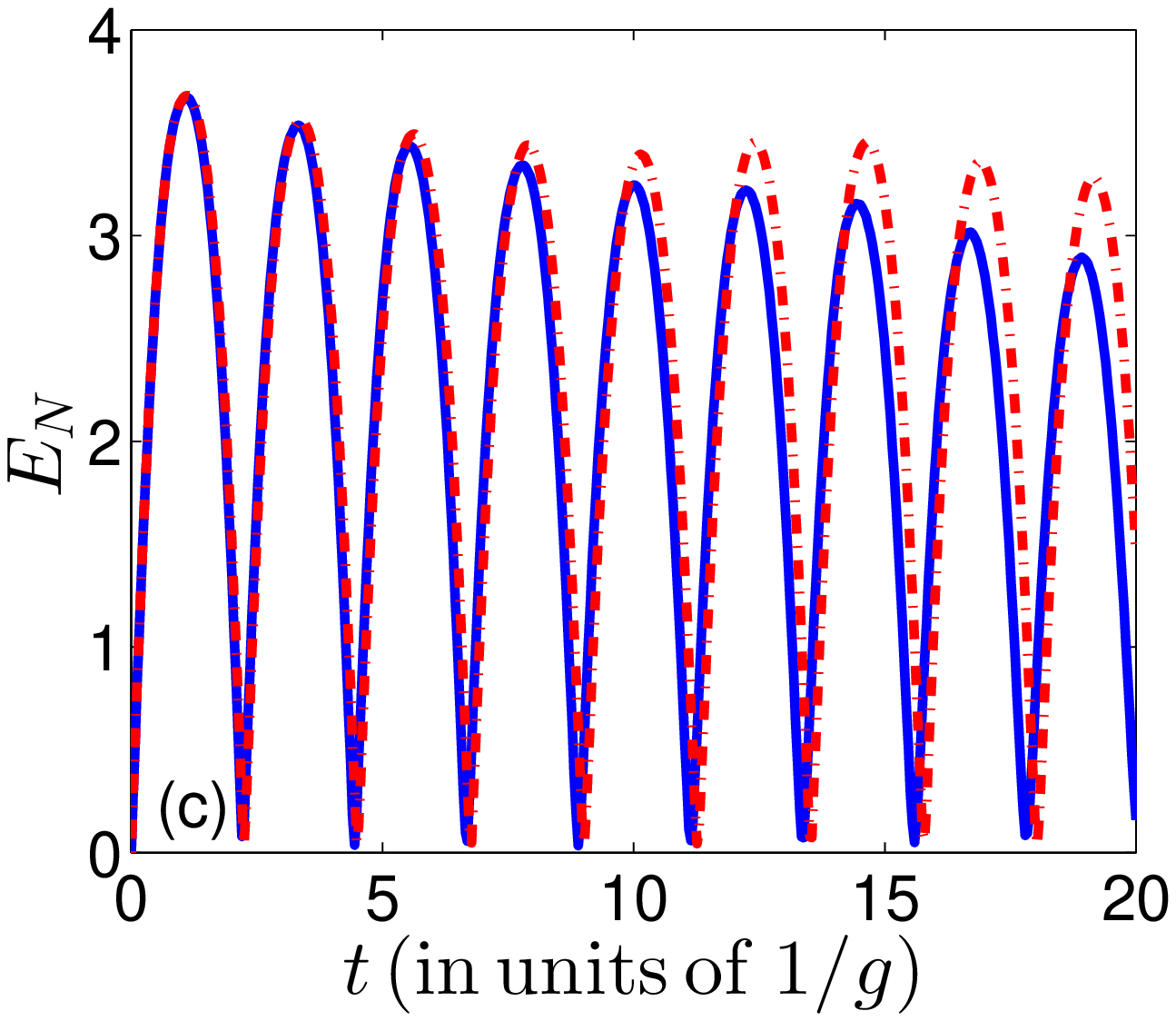}
\includegraphics[width=5.5cm]{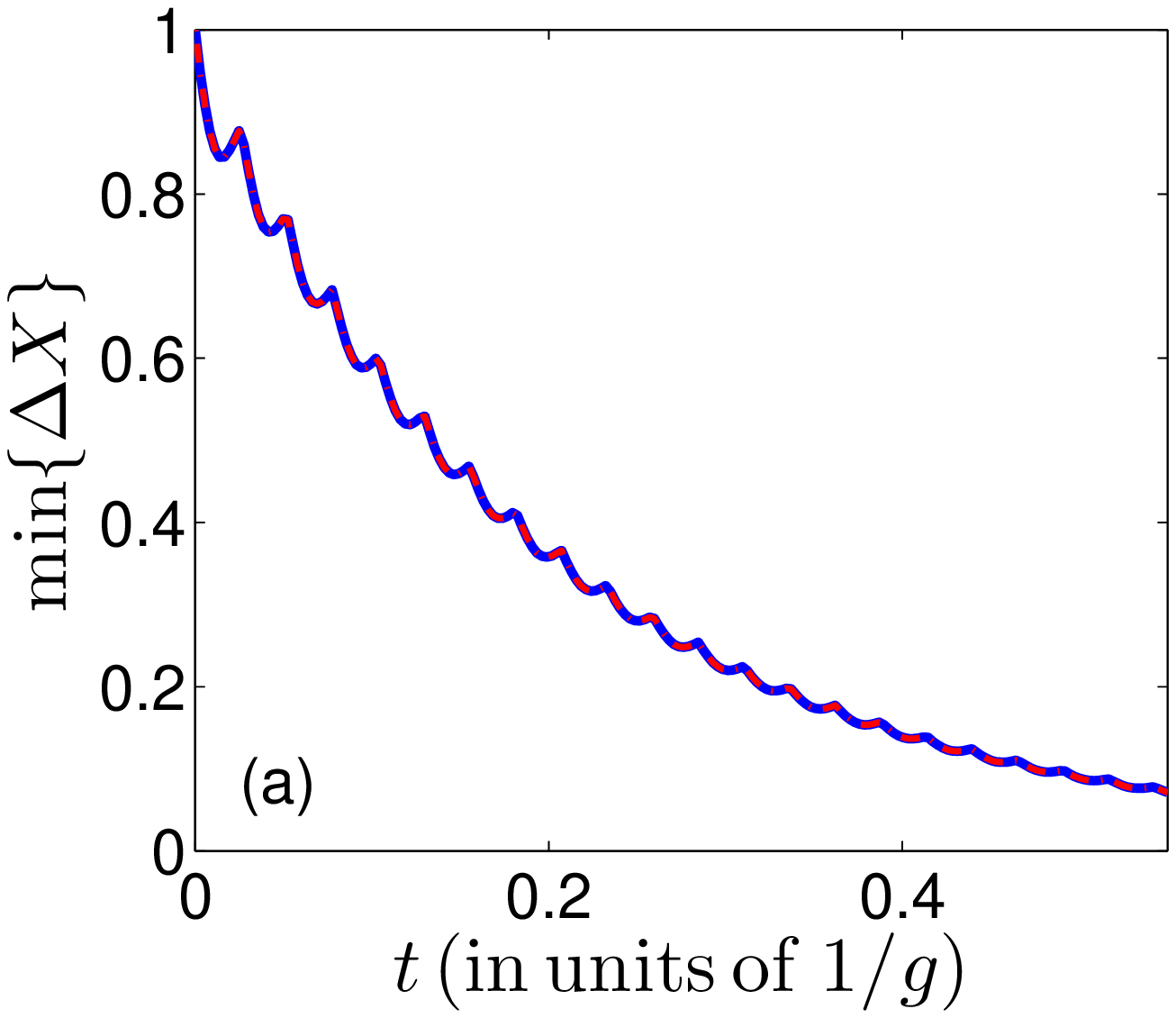}
\includegraphics[width=5.5cm]{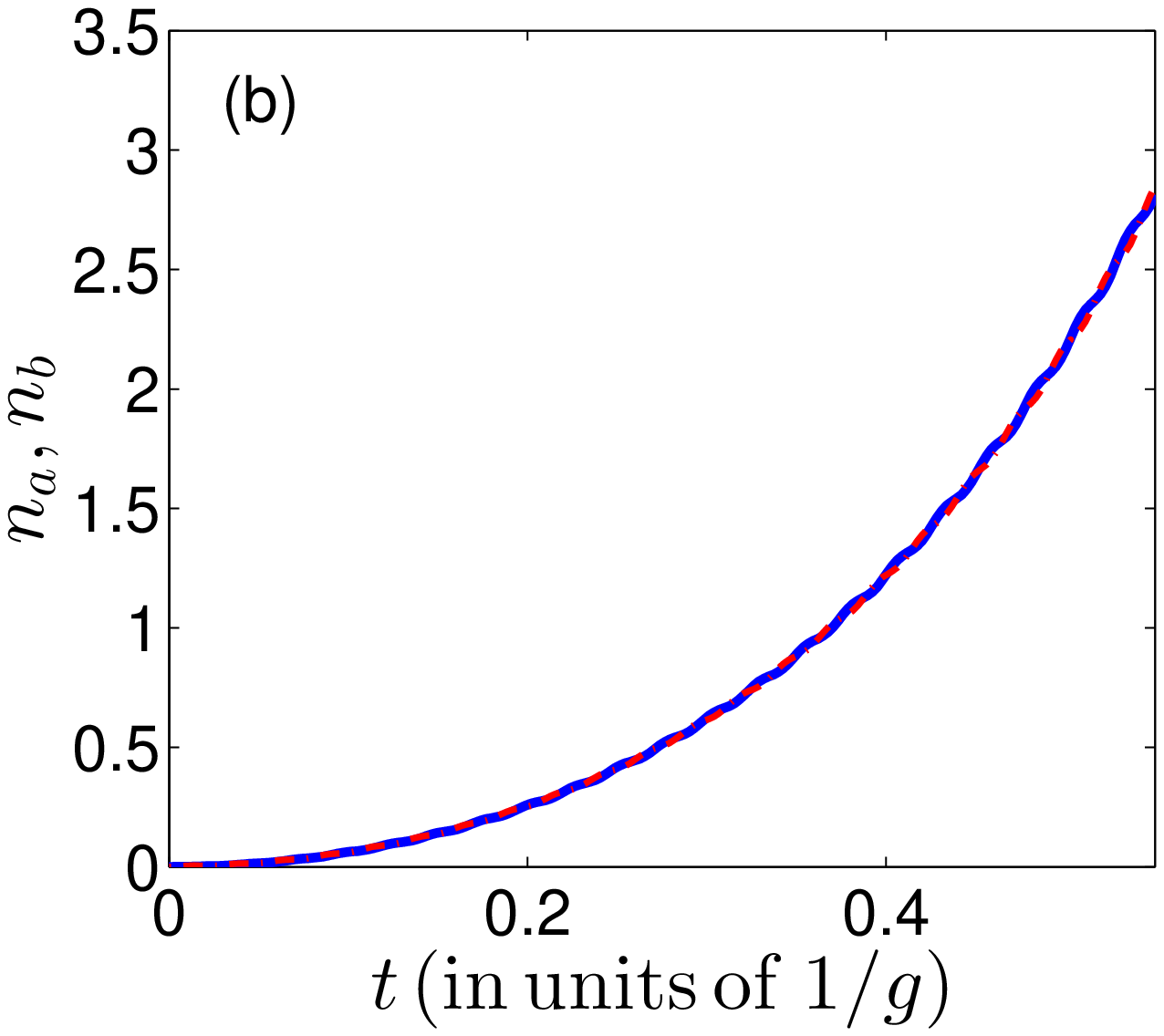}
\includegraphics[width=5.5cm]{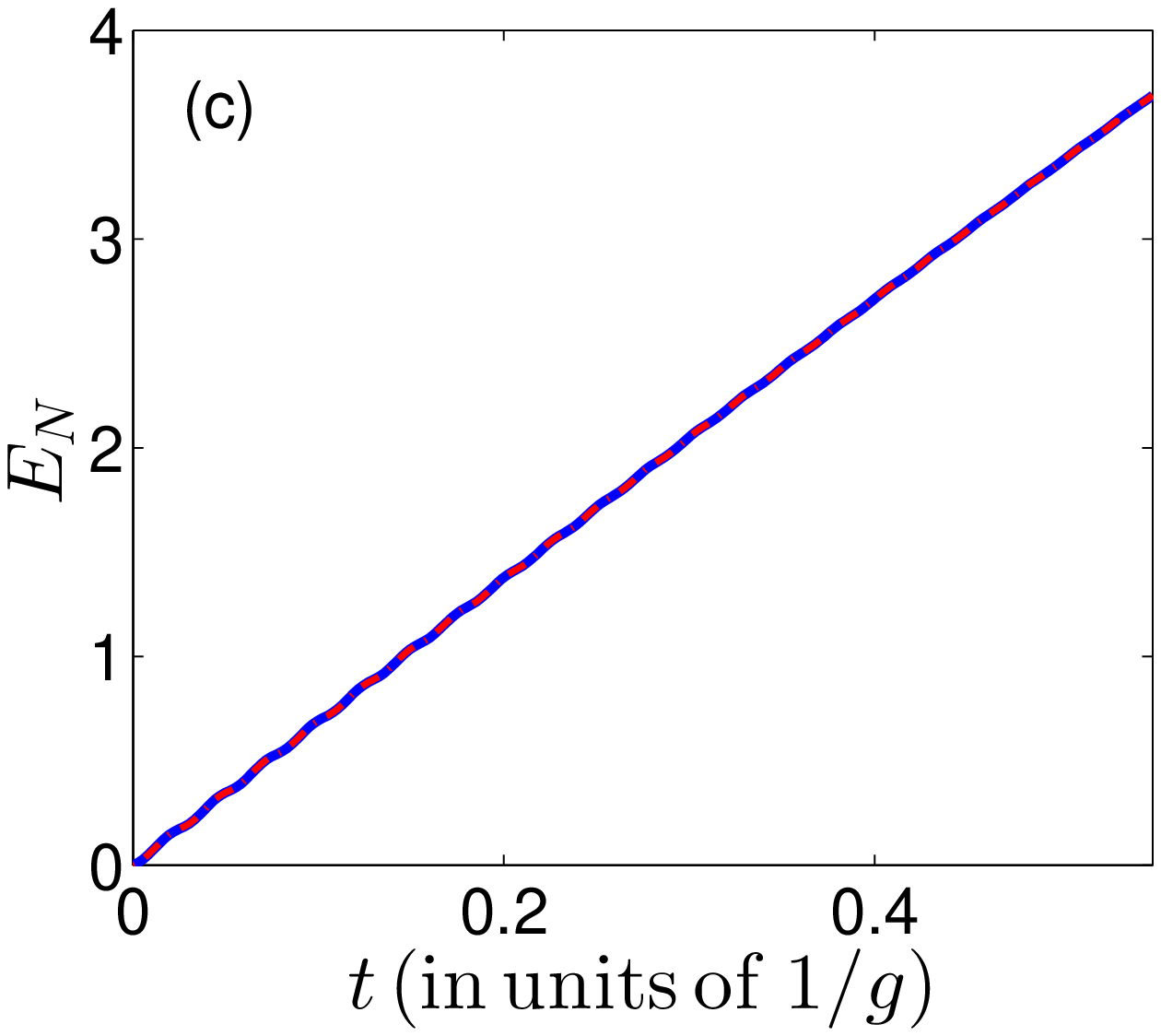}
\caption[]{(Color online) Same as in  Fig.~\ref{fig:2}, except for taking $N=1000$, $\Omega=30g$, $\delta_b=118g$, while $\delta_c$ is varied. In particular,  from top to bottom: $\delta_c=-125g$, $\delta_c=-123.3g$,
$\delta_c=-120g$.}\label{fig:3}
\end{figure*}

\section{Results}\label{Results}

In this section we study the system dynamics by numerical integration of master equation \eqref{ME} and we identify the regimes in which light and collective atomic modes are efficiently two-mode squeezed and entangled, assuming that the initial input state is the vacuum state for both cavity and spin-wave modes.

In order to gain insight into the dynamics, we first neglect dissipative processes and thus set $\gamma=\kappa=0$ in Eq.~\eqref{ME}. Our aim is to identify the regime of parameters for which the dynamics is Gaussian. This occurs when we can neglect the term scaled by the factor $\chi$, which represents a four-photon process involving three spin-wave and one cavity excitations. We thus compare the predictions of the dynamics governed by the full effective Hamiltonian $\eqref{Heff}$ (solid lines in the plots) with the ones found when the four-photon process in $H_{\rm eff}$ is discarded (dashed lines). In particular, we evaluate the time evolution of the logarithmic negativity $E_N$, Eq.~\eqref{EN}, the minimum of the composite quadrature as in Eq.~\eqref{Two:mode}, and the number of excitations per mode, namely,
\begin{eqnarray}
&&n_a(t)=\langle a\da(t)a(t)\rangle\,,\\
&&n_b(t)=\langle b_Q\da(t)b_Q(t)\rangle\,.
\end{eqnarray}
The number of excitations per mode, in particular, serves as a check of validity of the Holstein-Primakoff expansion, that is at the basis of our treatment.

Figure~\ref{fig:2} reports the time-dependent behaviour of the minimum variance of the composite quadrature, of the population of the field modes $n_a(t)$, $n_b(t)$, and of the logarithmic negativity $E_N(t)$, when dissipation is discarded and the detunings are so chosen that the dynamics is either periodic (upper row) or characterized by parametric amplification (lower row). In the first case the eigenvalues in Eq.~\rp{lambda} are purely imaginary ($\Lambda_2>0$ in Eq.~\eqref{Lambda:2}). When the four-photon processes scaled by $\chi$ in Eq.~\eqref{Heff} are neglected, the populations of the two modes (and correspondingly also two-mode squeezing and entanglement) oscillate in time. The choice of the parameter is such that the discrepancy between the results which do not include and which include the $\chi$ nonlinearity become evident as the occupation number per mode increases. On average, the effect of the four-photon-process term is to increase the minimum variance and to decrease the mean occupation per mode. This is easily understood since the $\chi$ nonlinearity scales with $n_b$ and makes the spectrum of excitations anharmonic, thus leading to a reduction in the absorption of laser photons. 

\begin{figure*}[t!]
\centering
\includegraphics[width=5.5cm]{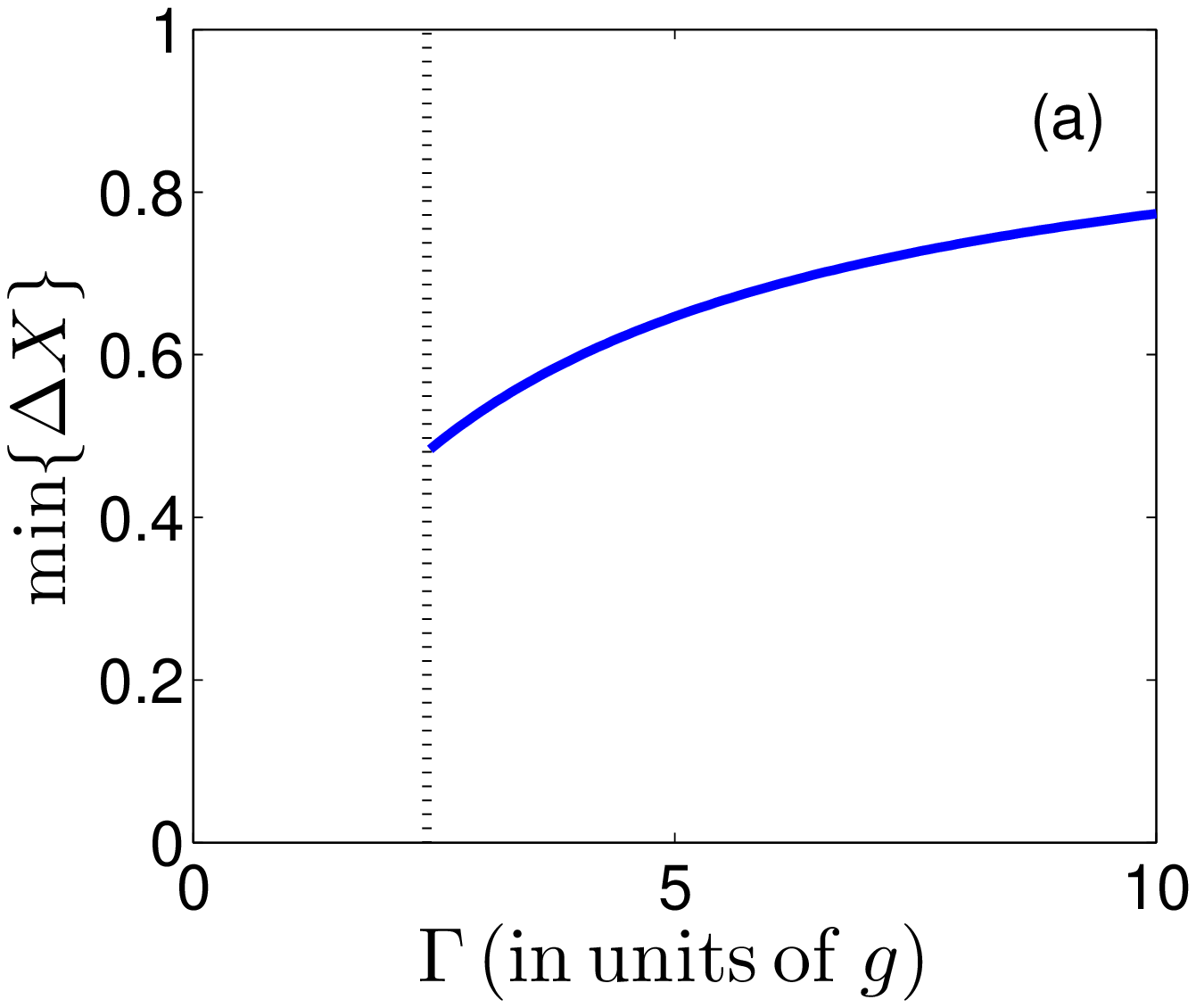}
\includegraphics[width=5.5cm]{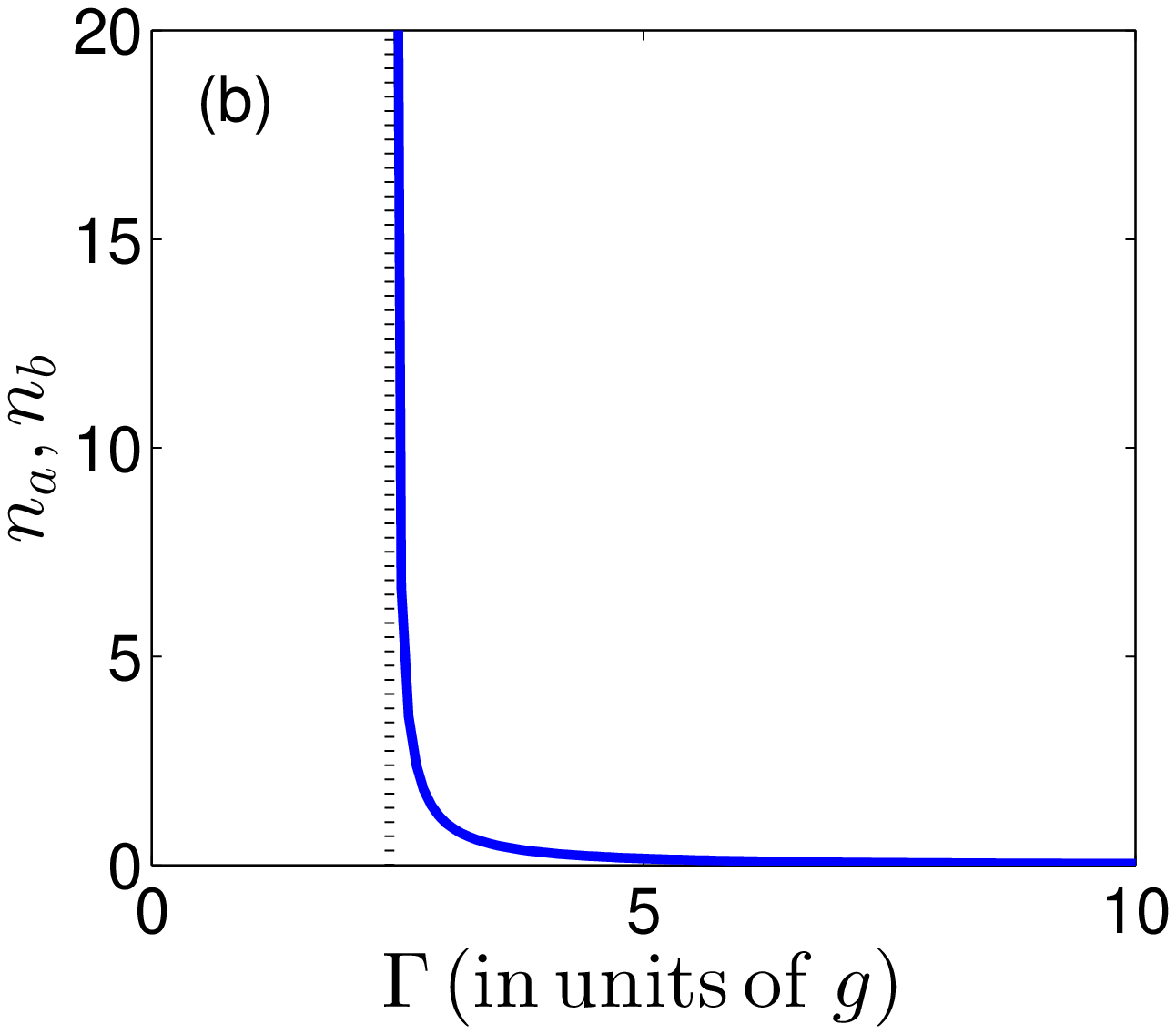}
\includegraphics[width=5.5cm]{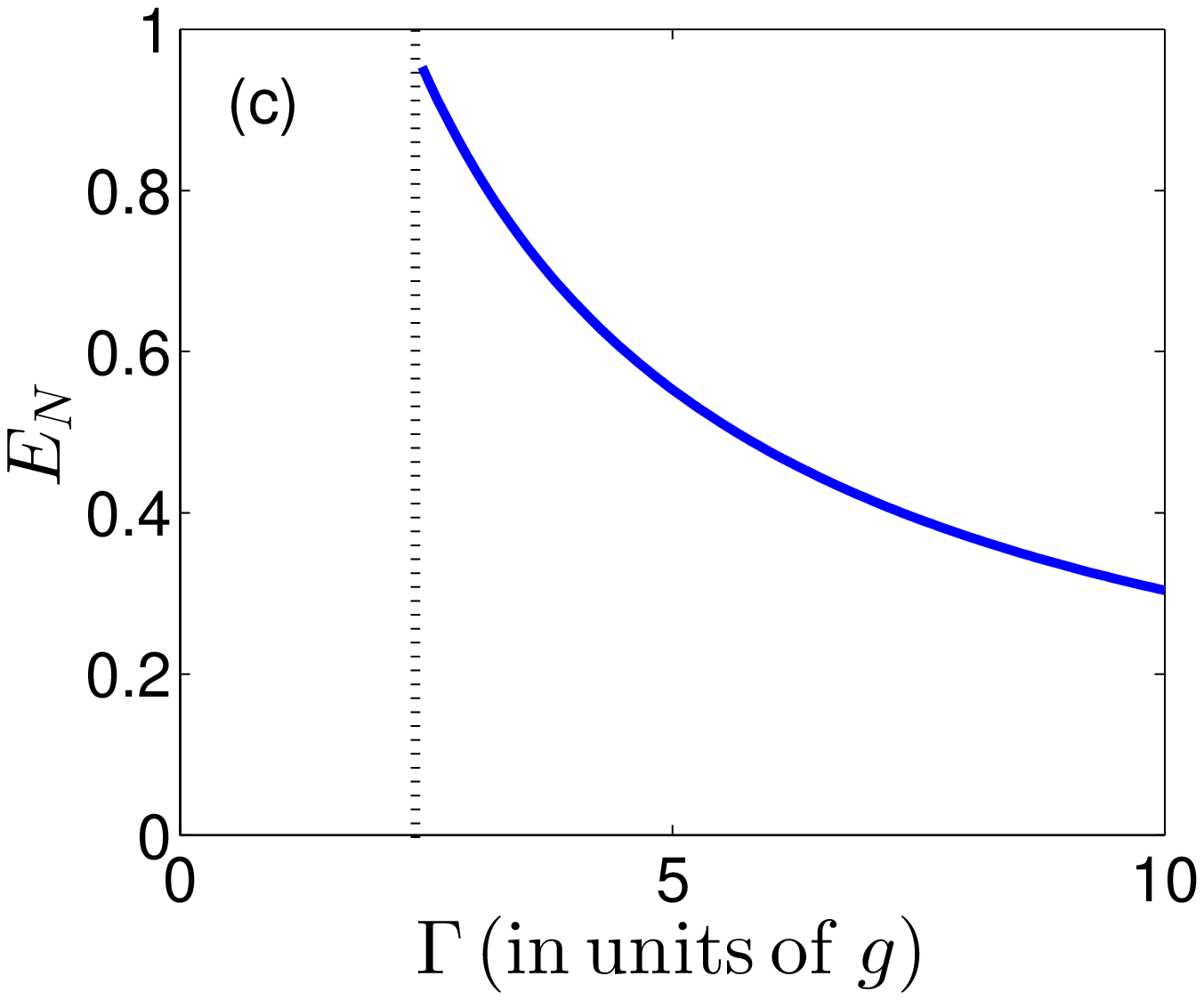}
\includegraphics[width=5.5cm]{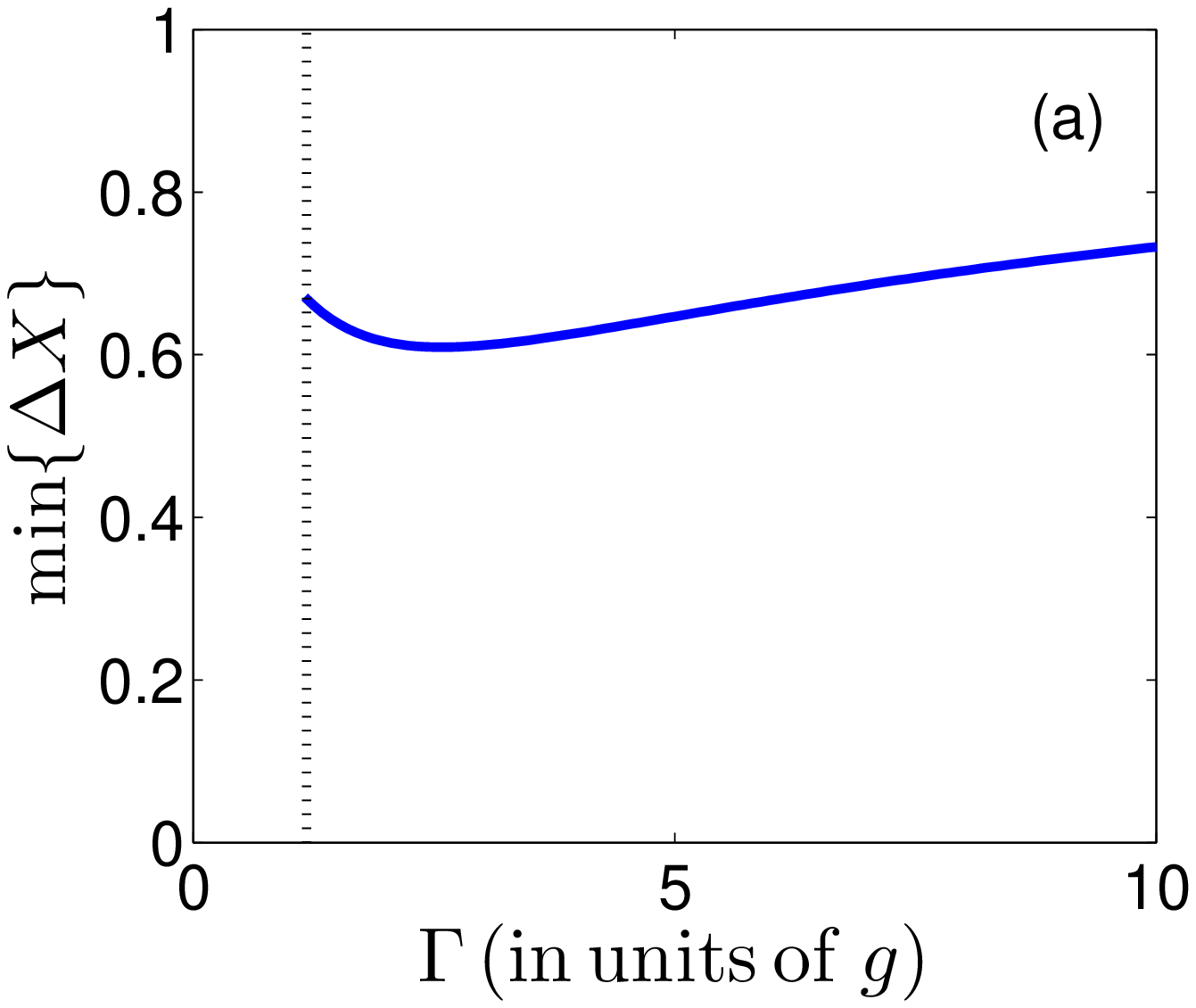}
\includegraphics[width=5.5cm]{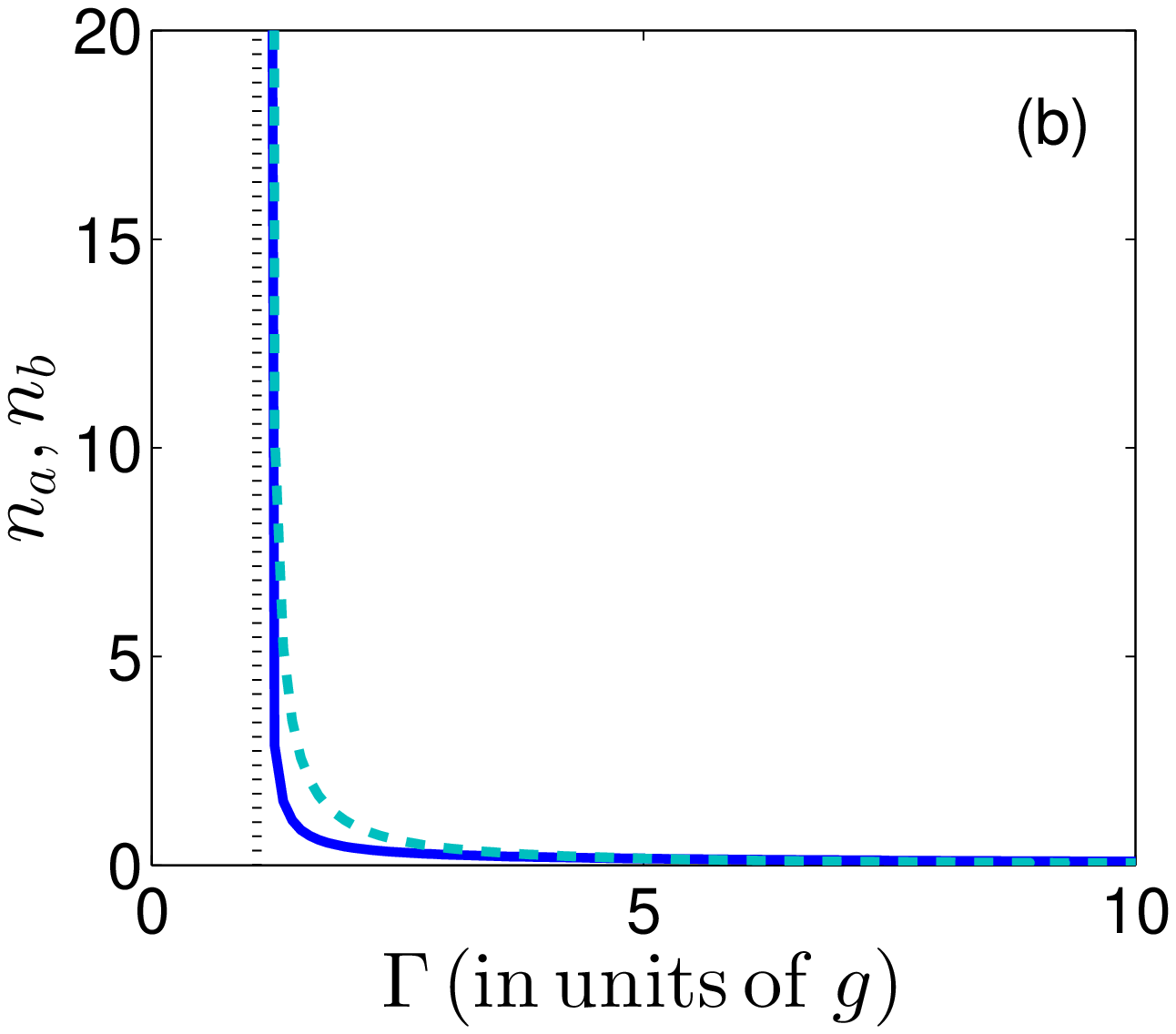}
\includegraphics[width=5.5cm]{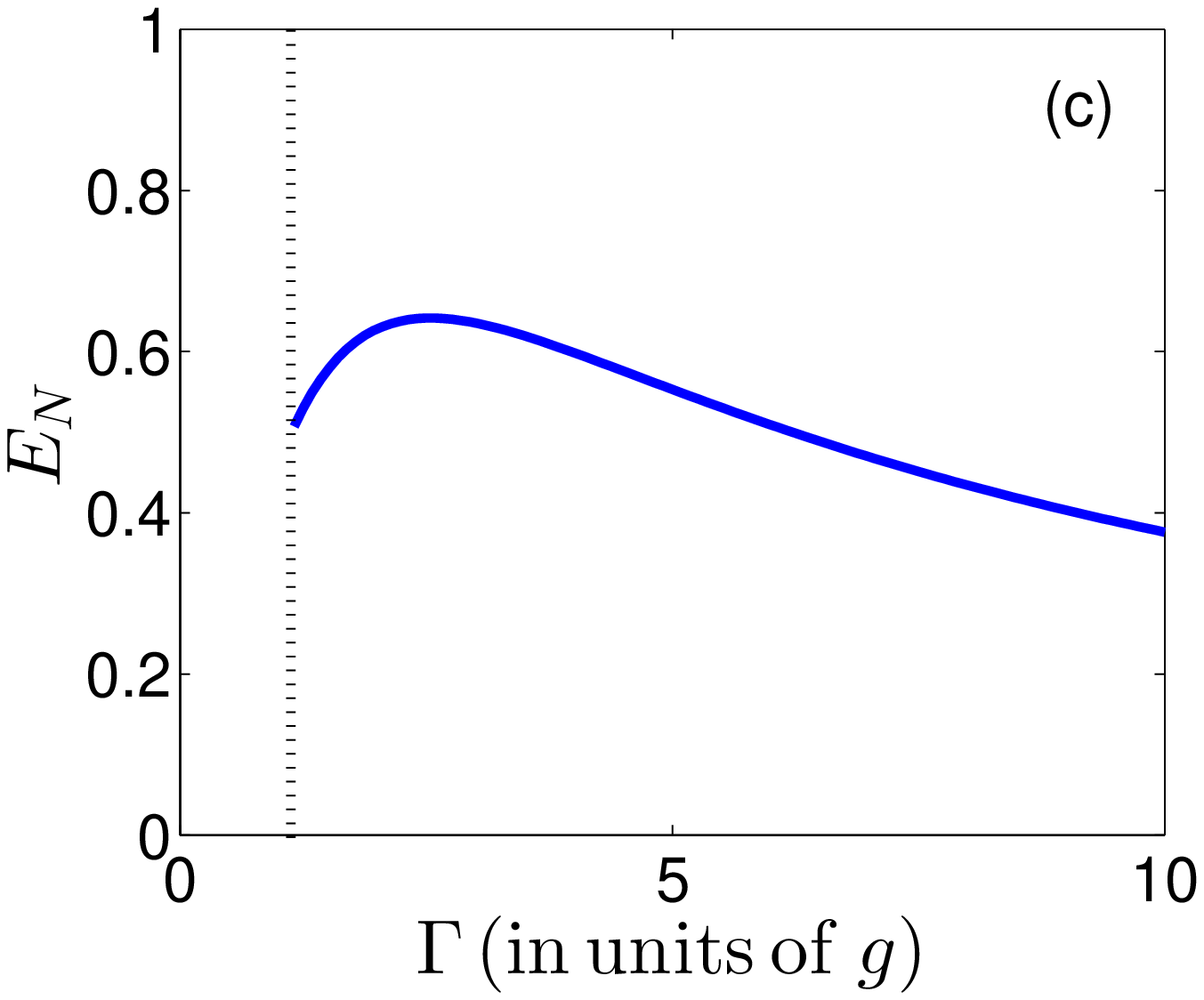}
\caption[]{(Color online) Stationary properties of the system in the presence of dissipation and when the dynamics gives rise to non-degenerate parametric amplification below threshold, as a function of the decay rate $\Gamma=\gamma/2$. The figures show (a) minimum of the variance of the composite quadrature, Eq.~\rp{Two:mode}, (b) the population of the field modes $n_a=\langle a^\dag a\rangle$ (solid blue line) and $n_b=\langle b_Q^\dag b_Q\rangle$ (dashed cyan line). Plot (c) corresponds to the Logarithmic negativity. The parameters of the Hamiltonian dynamics are as in Fig. \ref{fig:3}  with $\delta_c=-\delta_b=-118g$ and $\chi=0$, the dissipation rates are varied simultaneously, $\kappa=\gamma/2$ (top row), and $\gamma$ is varied while $\kappa=5g$ is kept fixed (bottom row). The quantities are determined using the steady state density matrix extracted from Eq.~\eqref{ME} (the vertical dotted lines indicate the threshold value, above which Master equation \eqref{ME} has a stationary solution).
}\label{fig:4}
\end{figure*}

The curves plotted in the lower row correspond to the case in which there is parametric amplification: The parameters are such that $\Lambda_2<0$ in Eq.~\eqref{Lambda:2}. The effect of the $\chi$-nonlinearity is visible after a transient time, in which the mode occupation $n_b$ has increased: It brings the minimum variance above the shot-noise limit, while it reduces the mode occupation, damping the energy stored in the system. The nonvanishing logarithmic negativity, in this case, indicates the presence of some entanglement in the system for certain time intervals, but the precise characterization and quantification of such entanglement is problematic, as the dynamics is now non-Gaussian and the simple theorems valid for the Gaussian case do not hold.

The relative weight of the non-Gaussian term is progressively reduced as the number of atoms and the Rabi frequency of the transverse laser are increased. Indeed, from Eqs.~\eqref{a:Qa} and \eqref{chi} we see that the weight of the scaling parameter $\alpha_{Q,a}$, that governs the term giving rise to two-mode squeezing, is increased. On the contrary, the nonlinearity $\chi\propto 1/\sqrt{N}$ and is thus suppressed as $N$ increases. The corresponding behaviors are reported in Fig.~\ref{fig:3}, where each row corresponds to different values of the detuning $\delta_c$, such that in the first two rows there is no amplification, being the eigenvalues imaginary ($\Lambda_2>0$), while in the bottom row the dynamics approaches that of a parametric amplifier. Comparison between the Gaussian and the full non-Gaussian dynamics shows that the $\chi$-nonlinearity is not relevant, even though small discrepancies are observed at sufficiently long times. The evolution in the regime of parametric amplification is analyzed over a reduced interval, due to the limited numerical capabilities in simulating the dynamics for large occupations. Over times of the order of $0.5g^{-1}$  we observe reduction of the minimum variance down to less than 10\% of the shot-noise limit, while the logarithmic negativity increases till $E_N\sim 4$. These results show that in a transient time, which can be shorter than the typical dissipation rates, spins and cavity mode can become entangled. A laser pulse driving the system over this time scale could thus lead to two-mode squeezing of the cavity and atomic degrees of freedom.

We now analyze the effect of dissipation and identify those regimes of parameters for which two-mode squeezing and entanglement are found in the asymptotic limit of the dynamics. The parameters we choose are those for which the dynamics can be considered Gaussian, paying attention to the fact that the parametric amplifier is below threshold: under this latter condition we can warrant that the approximations at the basis of our treatment are satisfied, so that the steady state we find is consistently described by master equation \eqref{ME}. The parameters for the coherent dynamics are here the same as in the bottom row of Fig.~\ref{fig:3}, where parametric amplification is found. We remark that the stationary state is determined after discarding the $\chi$ nonlinearity in  Hamiltonian \eqref{Heff}.

\begin{figure*}[t!]
\centering
\includegraphics[width=5.5cm]{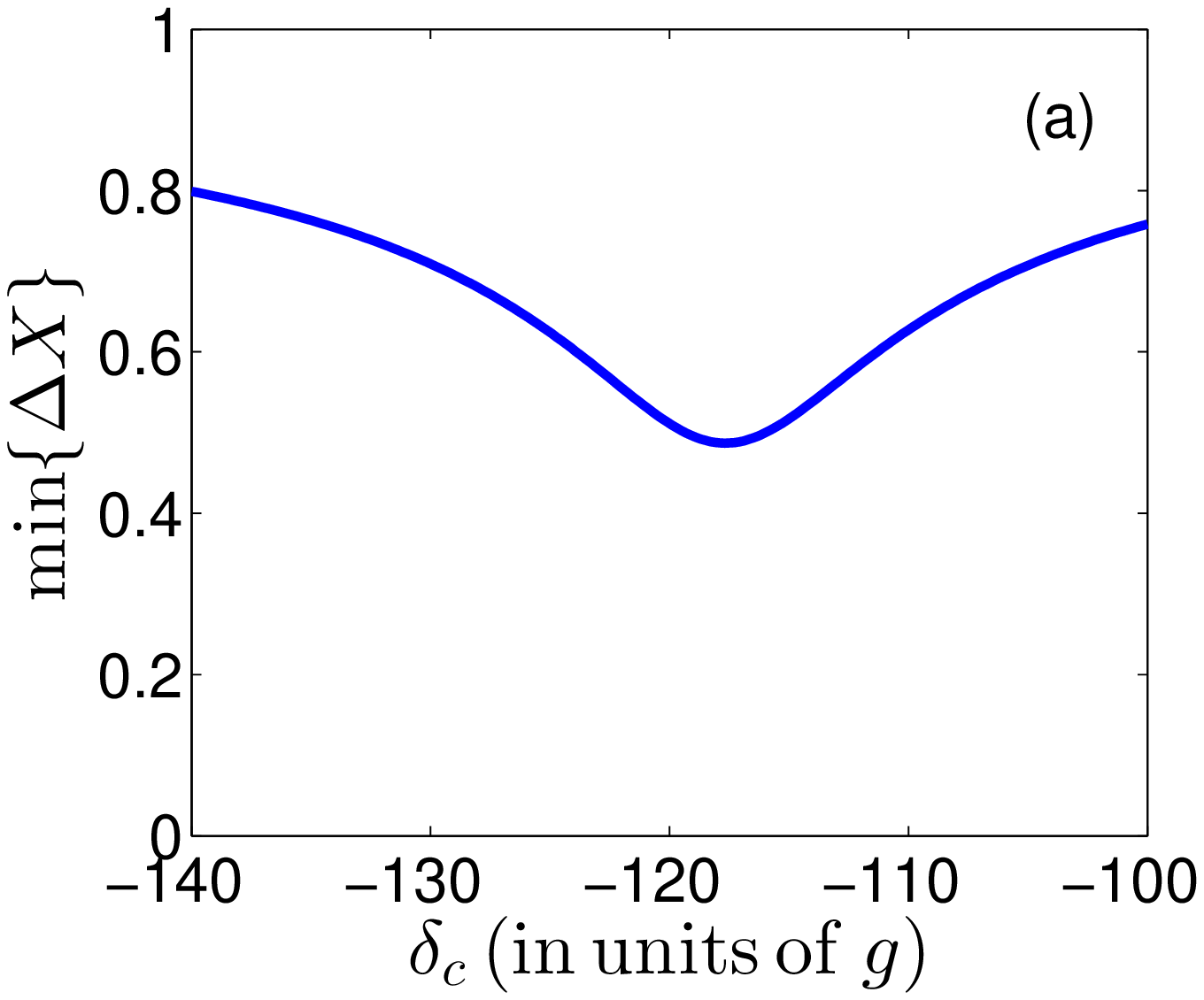}
\includegraphics[width=5.5cm]{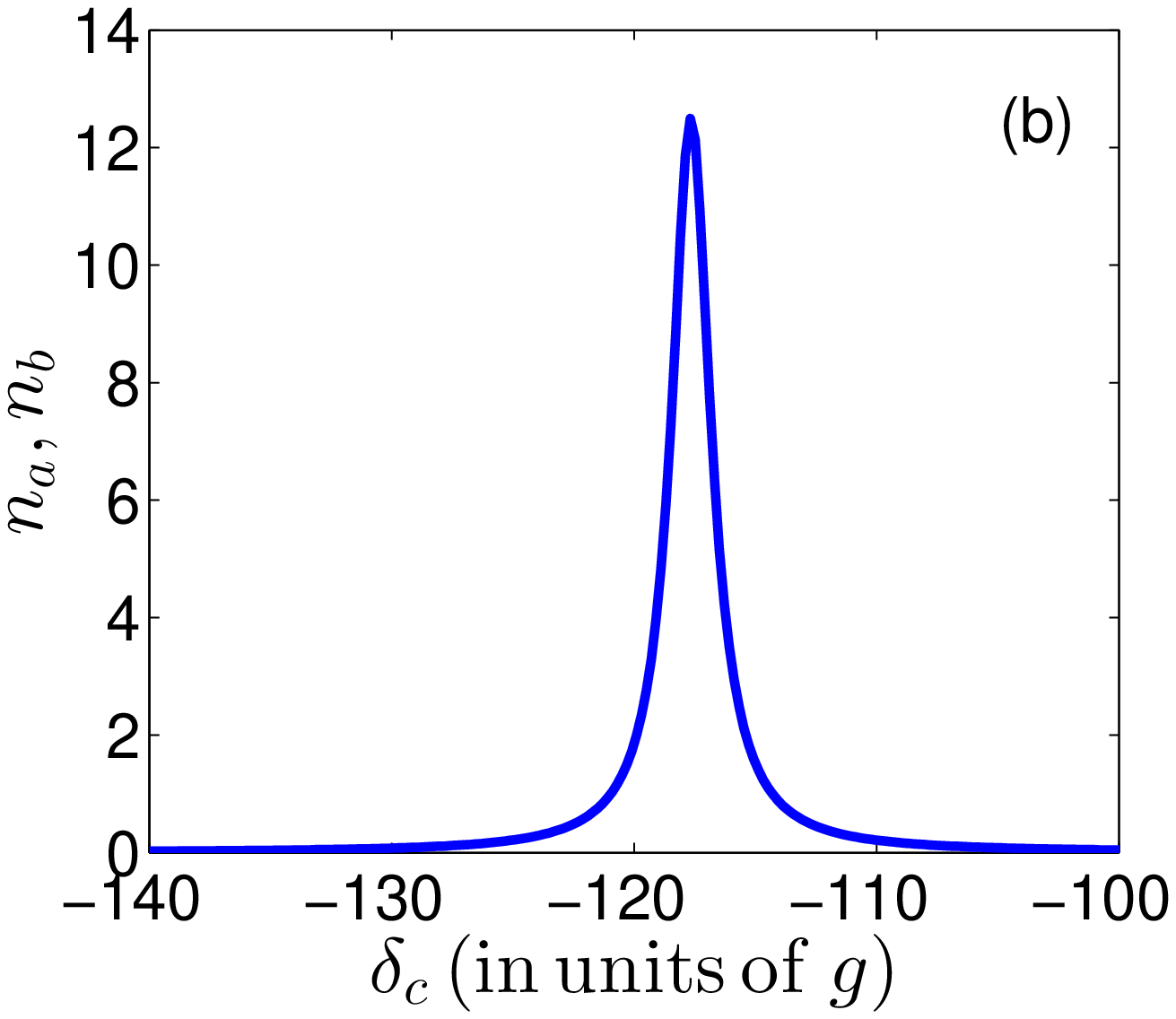}
\includegraphics[width=5.5cm]{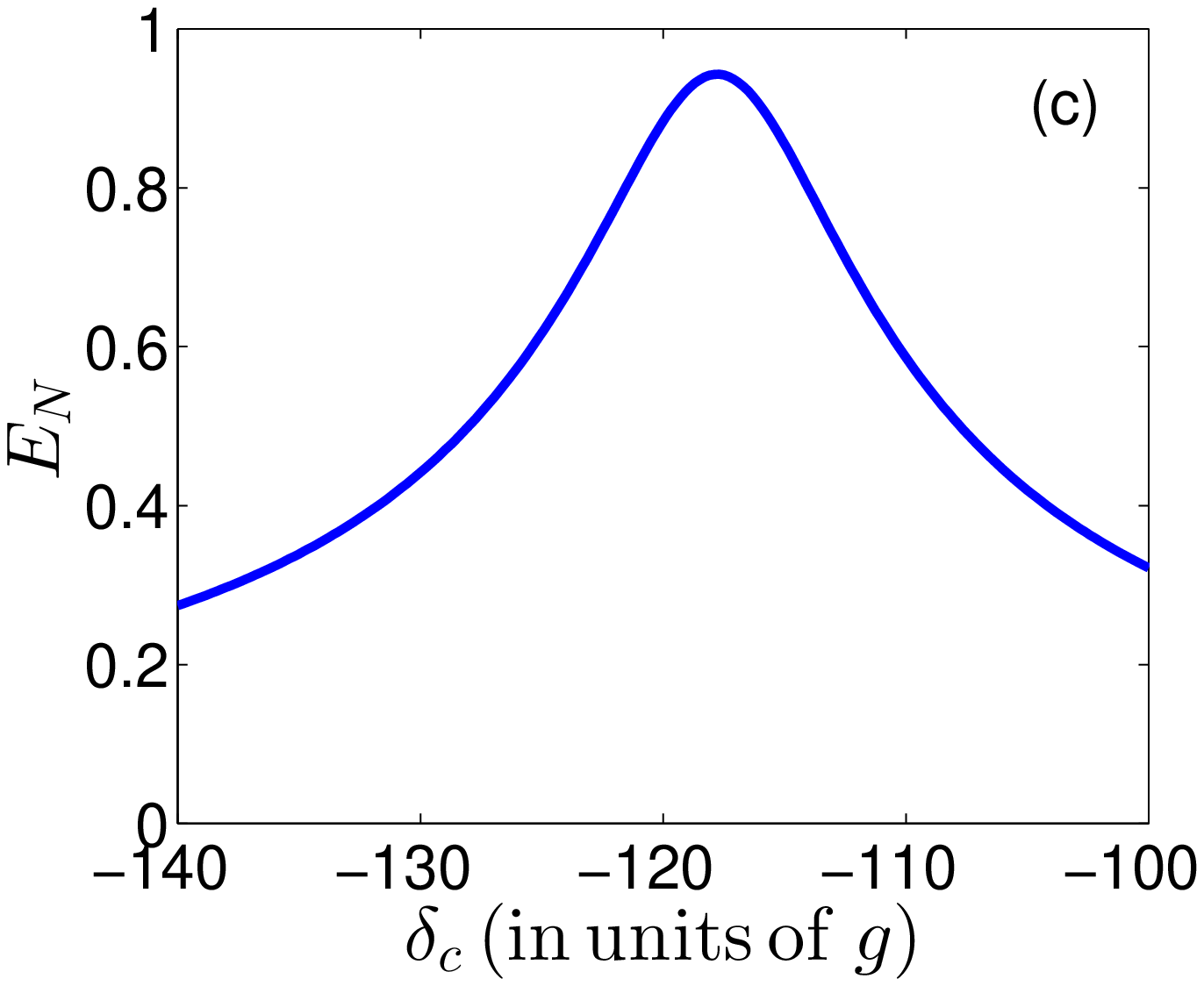}
\includegraphics[width=5.5cm]{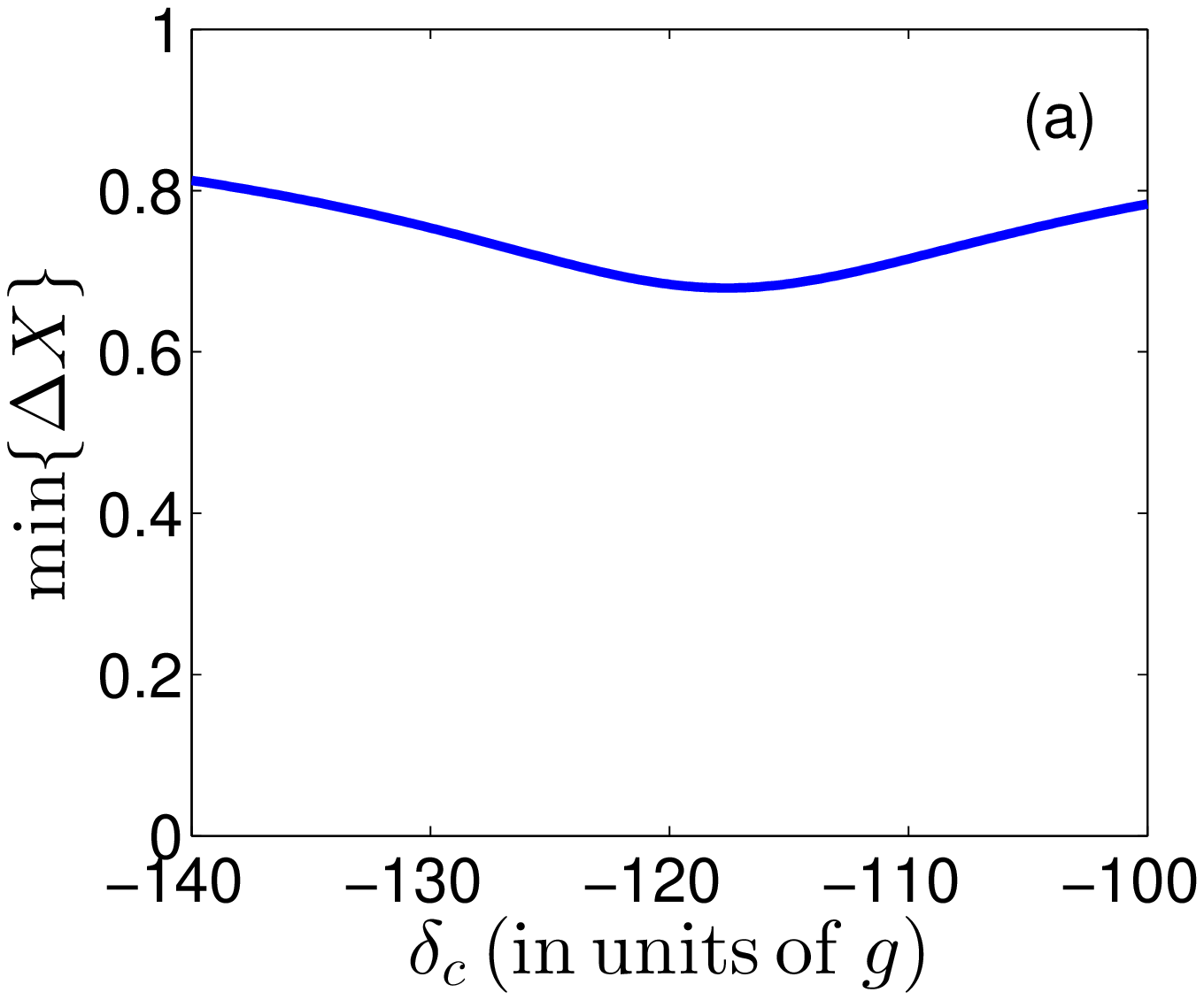}
\includegraphics[width=5.5cm]{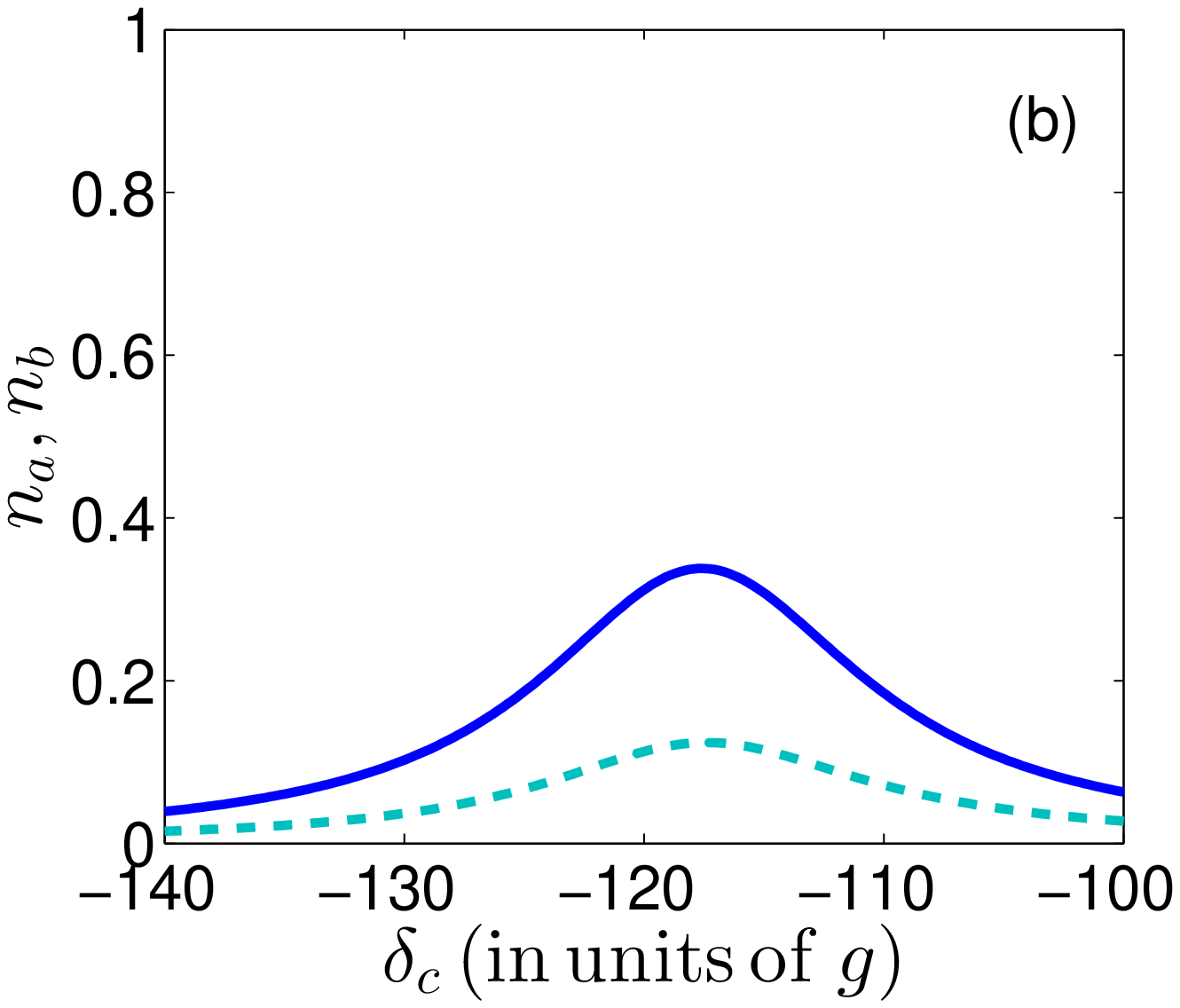}
\includegraphics[width=5.5cm]{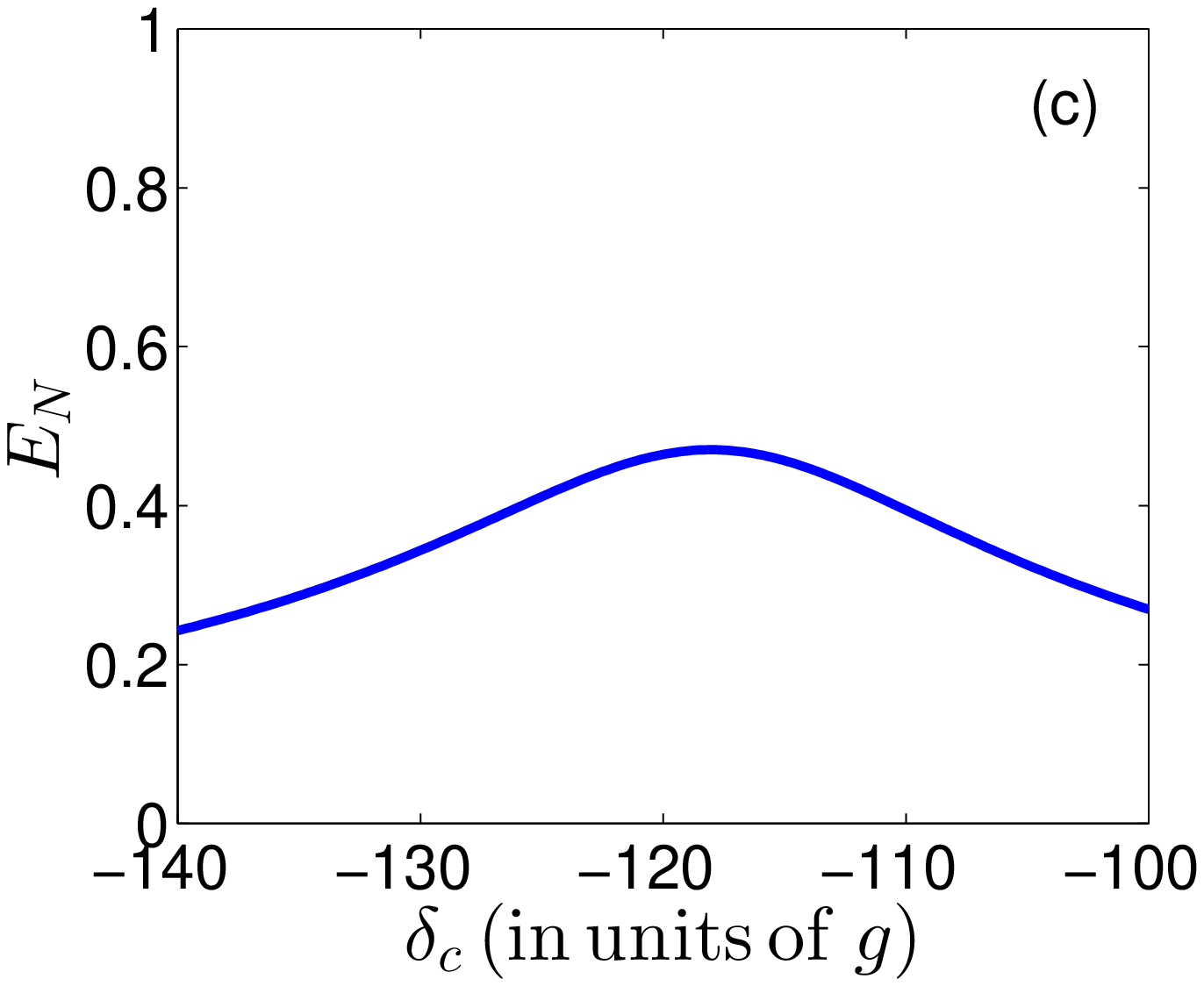}
\caption[]{(Color online) As in Fig.~\ref{fig:4}, where now the quantities are plotted as a function of $\delta_c$ for $\kappa=2.5g$ and $\gamma=5g$ (top row), $\gamma=15g$ (bottom row).
}
\label{fig:5}
\end{figure*}

Figure~\ref{fig:4} displays the variance minimum, the mean occupation per mode, and the logarithmic negativity at steady state as a function of the dissipation rates. In the top row we set $\kappa=\gamma/2$ and vary them simultaneously. Correspondingly, the single-atom cooperativity is  $C=(g/\kappa)^2$ and thus decreases with $\kappa^{-2}$ as $\kappa$ and $\gamma$ are increased. For comparison, the bottom row corresponds to the case in which only $\gamma$ is varied, while $\kappa$ is kept fix at the value $\kappa=5g$. The single-atom cooperativity in this case decreases with $C=g/(5\gamma/2)$ as $\gamma$ is increased.

We first discuss the plots in the upper row of Fig.~\ref{fig:4}, in which the decay rates are varied simultaneously. Here, we observe a monotonous behaviour as $\gamma$ and $\kappa$ are increased starting from the threshold value (indicated by the vertical line): The variance of the composite quadrature increases while entanglement decreases, showing that two-mode squeezing and entanglement are reduced because of dissipation. The occupation of both modes takes large values close to threshold, for which the validity of the Gaussian dynamics is questionable. In the bottom row, where only the effective radiative decay of the dipolar excitation is varied, the minimum variance exhibits a minimum away from threshold, where the occupation per mode is small. Correspondingly, the logarithmic negativity exhibits a maximum. This maximum occurs for values at which $\gamma \neq \kappa$, and shows that asymmetric decay rates can be advantageous for maximizing entanglement.  We note that, for $\gamma=2\kappa$ and close to threshold, the minimum variance is slightly smaller than the minimum value $0.5$ of a parametric amplifier below threshold~\cite{Reid1990}. This behaviour is due to the effect of the squeezing term  at strength $\alpha_{Q}$ in Hamiltonian (\ref{Heff}), as one can check by performing a numerical simulation in which this term is neglected.

Figure~\ref{fig:5} shows that two-mode squeezing is found when the detuning $\delta_c$ is equal to $-\delta_b$ for the cases $\gamma=2\kappa,6\kappa$. In both cases the minimum is, as expected, at the resonance $\delta_c=-\delta_b$. It is interesting to observe that the variance is below the shot-noise limit for a wide range of values, and correspondingly spins and light are entangled.

\begin{figure*}[t!]
\centering
\includegraphics[width=17cm]{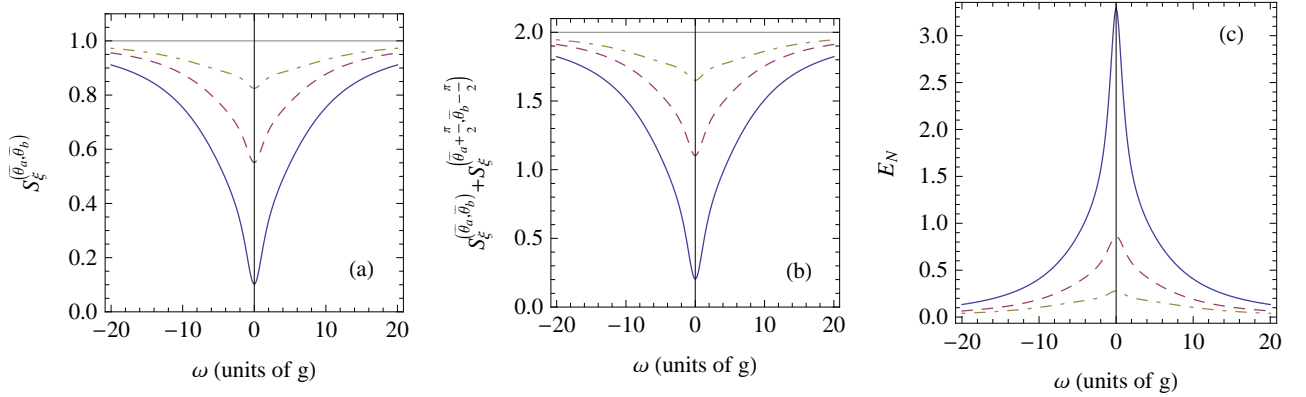}
\caption[]{(Color online) Stationary properties of the output fields collected at the detectors. (a) Two-mode squeezing spectrum of the emitted field as a function of the frequency (in units of $g$). The spectrum is evaluated from Eq.~\eqref{Somega} and reported for the optimal values of $\theta_a,\theta_b$ and $\xi$ which minimize the spectrum of squeezing and the entanglement criterion in Eq.~\eqref{S:sum} (the horizontal line indicate the shot noise level). (b) Corresponding entanglement criterion, Eq.~\eqref{S:sum}: when this value is smaller then 2 then the state is entangled. (c) Logarithmic negativity  $E_N$ of the state of the detected fields. The curves correspond to the collection efficiencies $\eta_a=\eta_b=1$ (blue solid lines), $\eta_a=\eta_b=0.5$ (red dashed lines). and $\eta_a=1$, $\eta_b=0.1$ (yellow dot-dashed lines). The parameters are $\kappa=2.5g$, $\gamma=15 g$, $\omega_z=100 g$ and $\delta_c=-\delta_b=-118 g$ and $\Omega=30 g$, $N=1000$; the local oscillators are tuned at the frequency of the normal modes $\Delta_\pm$, namely, $\Delta_a=-\Delta_b=-121.3g$.
}
\label{f1}
\end{figure*}

\begin{figure*}[t!]
\centering
\includegraphics[width=5.5cm]{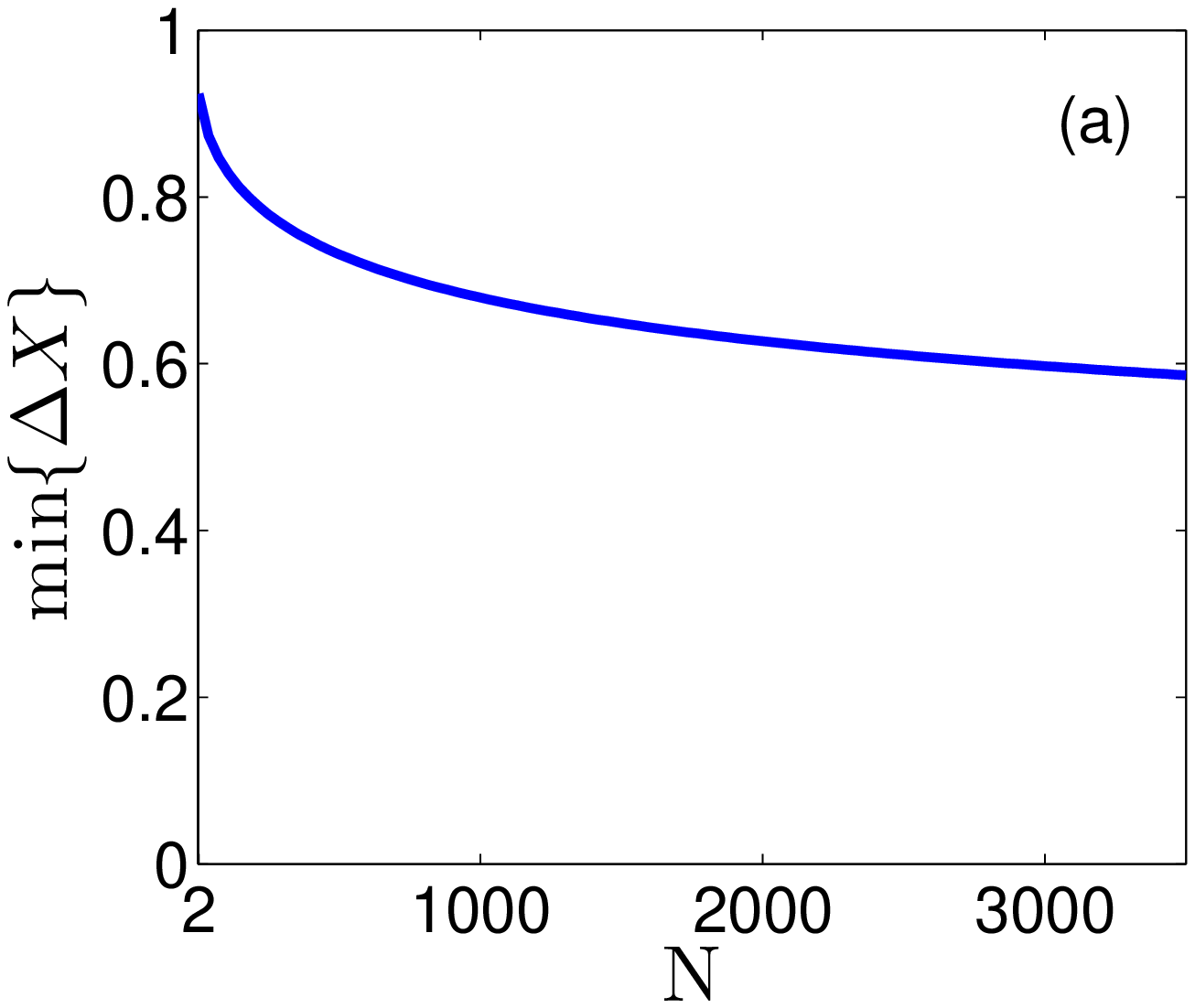}
\includegraphics[width=5.5cm]{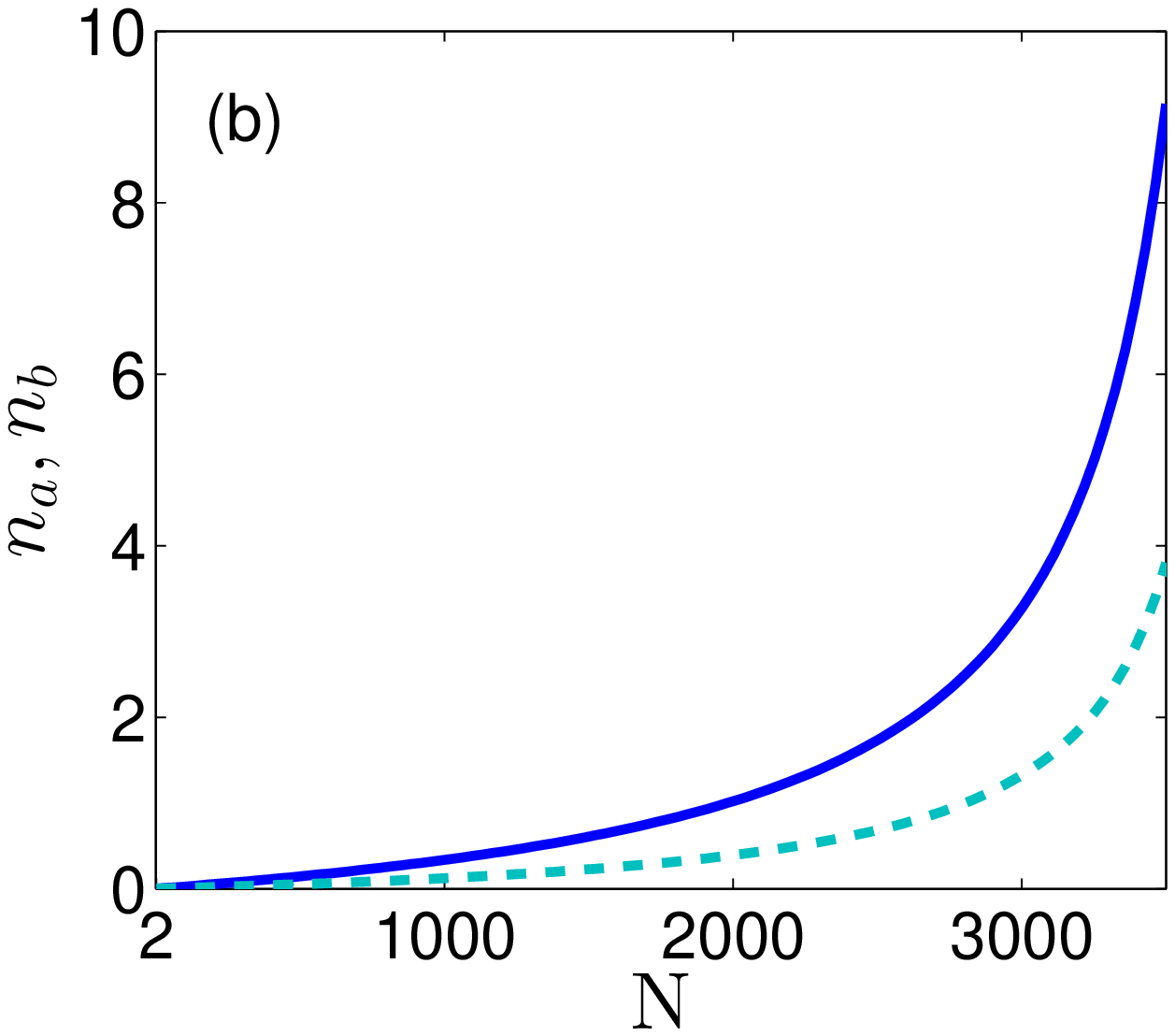}
\includegraphics[width=5.5cm]{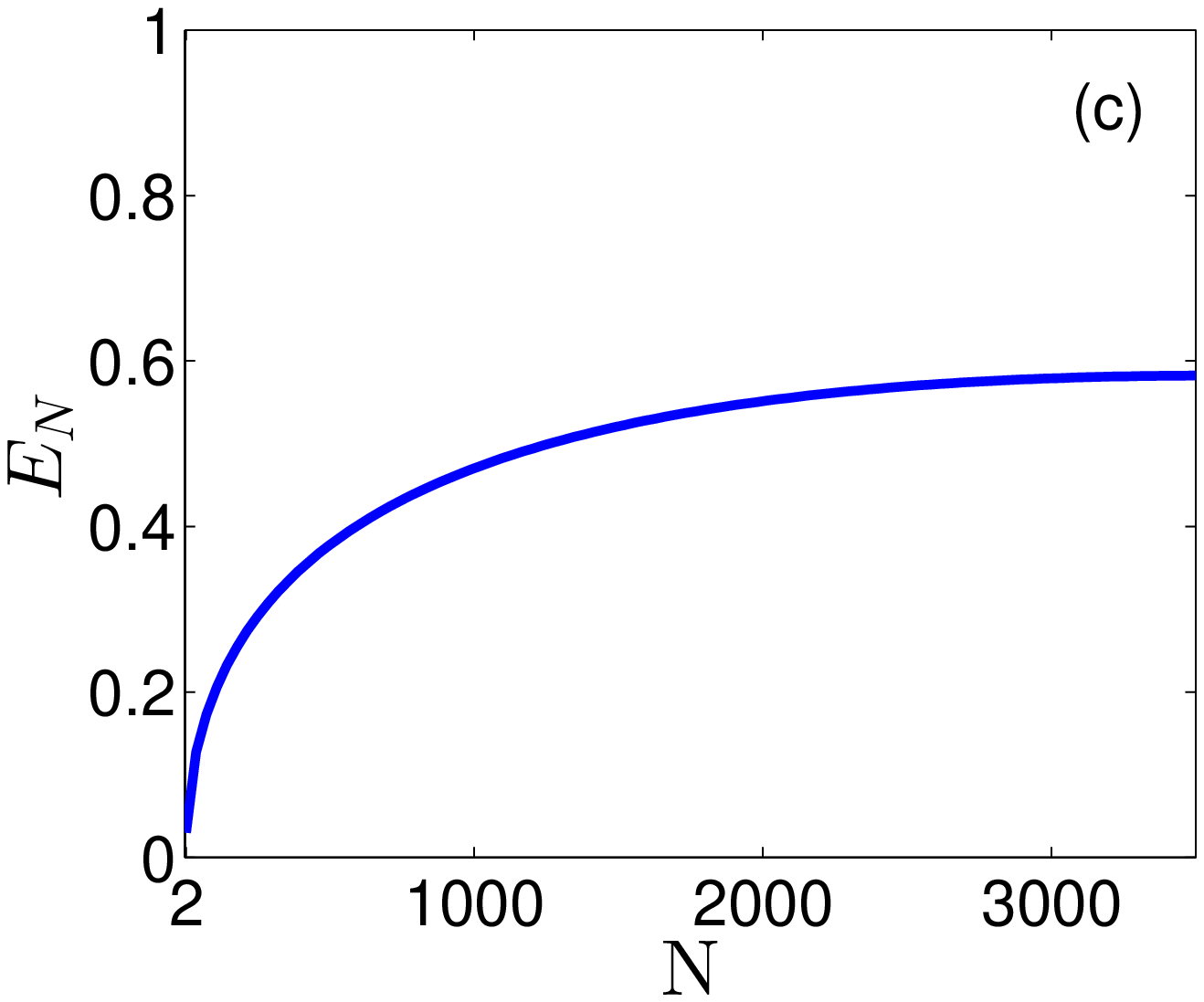}
\caption[]{(Color online) Results as in Fig.~\ref{fig:4}, as a function of the number of atoms $N$ for $\kappa=2.5g$ and $\gamma=15g$. In (b) the dark blue line represent $n_a$ and the dashed  cyan line $n_b$.}
\label{fig:6}
\end{figure*}

We now focus on the results for the regime of parameters investigated in recent cavity QED experiments~\cite{Vuletic}, where $(g,\kappa,\gamma/2)/2\pi\simeq(0.4,1,3)$ MHz, falling in the range of parameters considered in the bottom row of Fig.~\ref{fig:5}. For these parameters and $N=1000$ atoms we report the spectrum of squeezing in Fig. \ref{f1}(a) for different detectors collection efficiencies. The corresponding condition of entanglement is displayed in subplot (b), while the logarithmic negativity of the covariance matrix is reported in (c). This plot shows that for finite collection efficiency some entanglement is still present in the emitted fields, the reduction is determined by the collection efficiency. We remark that in these figures we have optimized the values of $\theta_a,\theta_b,\xi$ and refer the reader to the appendix for further details.

Finally, we analyze the entanglement as a function of the cooperativity, which increases linearly with number of atoms. Figure~\ref{fig:6} shows the behaviour of minimum variance of the composite quadrature, occupation of the modes, and logarithmic negativity as the number of atoms $N$ is increased (corresponding to a linear increase of the $N$-atom cooperativity). Two-mode squeezing and entanglement increase with $N$ but seem to saturate, while the occupation of the modes increase exponentially. We note that, the larger is $N$, the better the dynamics is approximated by a Gaussian model. In other words, non-Gaussian effects become evident at small atom numbers.

\section{Conclusions}\label{Conclusions}

A collective spin mode of an array of atoms, strongly coupled to the mode of an optical cavity and driven by a transverse laser, can become entangled with the cavity field. This entanglement is robust against dissipation and is present at the steady-state of the dynamics. It relies on a phase-matching condition between the cavity wavelength and the inter-particle spacing within the atomic array. We have restricted our analysis to the limit in which nonlinear contributions, giving rise to non-Gaussian elements of the dynamics, can be neglected. In this limit we could determine analytically the spectrum of the squeezing and put it in a precise relation to the entanglement by exploiting the exact general results holding for the logarithmic negativity in the case of Gaussian states. In this way we were able to quantify the amount of entanglement generated between spins and cavity field. Non-Gaussianity, on the other hand, can be an important resource for quantum information protocols with continuous variables, especially for tasks, like entanglement distillation and measurement-based quantum computation, that cannot be achieved within a purely Gaussian framework. Indeed, the system that we have considered can be tuned well into the non-Gaussian regime by suitably choices of the coupling parameters. In this case, the entanglement which is generated can be quantified by means of multiple quantities, including operational ones such as for instance the teleportation fidelity~\cite{Pirandola03,Pirandola06,Dellanno2007,Rebic2010,Dellanno2013}, and geometric ones such as the degree of non-Gaussianity as measured by the distance or the relative entropy from a reference Gaussian state~\cite{Genoni}, and the affinity of a non-Gaussian resource state with respect to a reference two-mode squeezed vacuum~\cite{Dellanno2010,Dellanno2007}.

The results that we have obtained correspond to experimentally accessible parameters~\cite{Vuletic} and in agreement with the reduction of quadrature fluctuations at steady state, as predicted for hot atomic ensembles and optical depths of the order of 30~\cite{Muschik}. Therefore, one can consider atomic arrays in resonators as valid platforms for the realization of steady-state entanglement between collective atomic spins and light, constituting a tunable alternative to hot atomic ensembles in cells~\cite{Muschik,Krauter}. We further observe that in our analysis we neglected the motion of the atoms inside the resonator, assuming that the atoms are tightly trapped in the minima of the optical lattice. The coupling with these degrees of freedom has been analyzed in a similar setting as a function of the spatial localization~\cite{FernandezVidal}. At the same time, it can be considered a further degree of freedom which can be interfaced with light, in the spirit of various previous proposals~\cite{ParkinsKimble,MorigiPRL2006}.

Finally,  it would be interesting to extend the present scheme by replacing the laser with an incoherent, nonclassical and entangled, squeezed bath. Recent works on similar setups, in fact, predict maximal entanglement transfer between qubits and continuous variables as well as indefinite, scale-free replication of the entanglement present in the nonclassical driving field~\cite{Campbell2010,Zippilli2013}.

\begin{acknowledgments}
The authors are grateful to Geza Giedke and Oxana Mishina for helpful comments. We acknowledge financial support from the European Commission of the European Union under the FP7 STREP Project iQIT (integrated Quantum Information Technologies), Grant Agreement n. 270843; the FP7 STREP Project EQuaM (Emulators of Quantum Frustrated Magnetism), Grant Agreement n. 323714; the FP7 Integrated Project AQUTE; and the FP7
STREP Project PICC; by the Spanish Ministerio de Ciencia y Innovaci{\'o}n (QOIT: Consolider-Ingenio 2010; QNLP: FIS2007-66944; FPI), by BMBF (QuORep, Contract No. 16BQ1011), and by the German Research Foundation.
\end{acknowledgments}

\appendix*

\section{Evaluation of the spectrum of squeezing and of the logarithmic negativity of the spectral components of the detected field}

For calculating the spectrum of squeezing we consider the equations of motion in Fourier space. Let $\tilde{\zeta}(\omega)$ be the Fourier transform at frequency $\omega$ of the field $\zeta(t)=a,b_Q,a\da,b_Q\da$, with
$$\tilde{\zeta}(\omega)=\frac{1}{\sqrt{2\pi}}\int\dd t \ee^{\ii\omega t}\zeta(t)\,.$$
We first note that, according to this definition, $[\tilde{\zeta}(\omega)]\da=\tilde{\zeta}\da(-\omega)$. The corresponding vector of operators is related to the vector $\va(t)$ by the Fourier transform
$$\tilde{\va}(\omega)=\frac{1}{\sqrt{2\pi}}\int\dd t \ee^{\ii\omega t}\va(t)\,.$$
We further denote by $\tilde{\va}_{\ell}^{\rm in}(\omega)$ and $\tilde{\va}_{\ell}^{\rm out}(\omega)$, with $\ell \in \pg{d, nd}$, the Fourier transform of $\va_\ell^{\rm in}(t)=\pt{a_{\rm in}^{(\ell)}(t),b_{\rm in}^{(\ell)}(t),a_{\rm in}^{\dagger(\ell)}(t),b_{\rm in}^{\dagger(\ell)}(t)}$ and $\va_\ell^{\rm out}(t)=\pt{a_{\rm out}^{(\ell)}(t),b_{\rm out}^{(\ell)}(t),a_{\rm out}^{\dagger(\ell)}(t),b_{\rm out}^{\dagger(\ell)}(t)}$, respectively.
The Heisenberg-Langevin equations~\rp{HL:a:nd} in Fourier space read
\begin{eqnarray}\label{aomega1}
\tilde \va^{\rm out}_d(\omega)&=&\QQ'\,\tilde \va(\omega)-\tilde \va_d^{\rm in}(\omega),\\
\label{aomega2}
-\ii\omega\,\tilde \va(\omega)&=&\ZZ\,\tilde \va(\omega)+\QQ'\,\tilde \va_d^{\rm in}(\omega)+\QQ''\,\tilde \va_{nd}^{\rm in}(\omega),
\end{eqnarray}
where the matrices $\QQ'$ and $\QQ''$ are diagonal and their diagonal elements are $\pt{\sqrt{2\eta_a\kappa},\sqrt{\eta_b\gamma},\sqrt{2\eta_a\kappa},\sqrt{\eta_b\gamma}}$, and $\pt{\sqrt{2(1-\eta_a)\kappa},\sqrt{(1-\eta_b)\gamma},\sqrt{2(1-\eta_a)\kappa},\sqrt{(1-\eta_b)\gamma}}$ respectively.

Using Eq. \eqref{aomega2} in Eq. \eqref{aomega1} one finds a linear equation connecting output and input fields,
$\tilde \va_d^{\rm out}(\omega)=-\WW'(\omega)\,\tilde \va_d^{\rm in}(\omega)-\WW''(\omega)\,\tilde \va_{nd}^{\rm in}(\omega),$
where
\begin{eqnarray}
\WW'(\omega)&=&\QQ'\pt{\ZZ+\ii\omega}^{-1}\QQ'+\id\,,
\nn\\
\WW''(\omega)&=&\QQ'\pt{\ZZ+\ii\omega}^{-1}\QQ'' \,.
\end{eqnarray}
The spectrum of the correlation matrix for the output field, whose elements are 
$\widetilde{\AAA}_{j,k}(\omega,\omega')=\av{\pg{\tilde{\va}_d^{\rm out}(\omega)}_j\pg{\tilde{\va}_d^{\rm out}(\omega')}_k}$, 
is
\begin{eqnarray}\label{AAA}
\widetilde{\AAA}(\omega,\omega')&=&
\av{\tilde{\va}_d^{\rm out}(\omega)\ \tilde{\va}_d^{\rm out}(\omega')^T}
\nn\\
&=&\delta(\omega+\omega')\widetilde{\AAA}_0(\omega)
\end{eqnarray}
with
 \begin{eqnarray}\label{AA0}
\widetilde \AAA_0(\omega)=\WW'(\omega)\ \GG\ \WW'^T(-\omega)+\WW''(\omega)\ \GG\ \WW''^T(-\omega)\,,
\end{eqnarray}
where we have used the fact that the modes $\tilde \va_{d}^{\rm in}(\omega)$ and $\tilde \va_{nd}^{\rm in}(\omega)$ are decorrelated, and the matrix $\GG$ is defined as
\begin{eqnarray}
{\cal G}=\left(
\begin{array}{cc}
 0 & \openone  \\
 0 & 0  \\
\end{array}
\right)\,.
\end{eqnarray}

\begin{figure*}[t!]
\centering
\includegraphics[width=17cm]{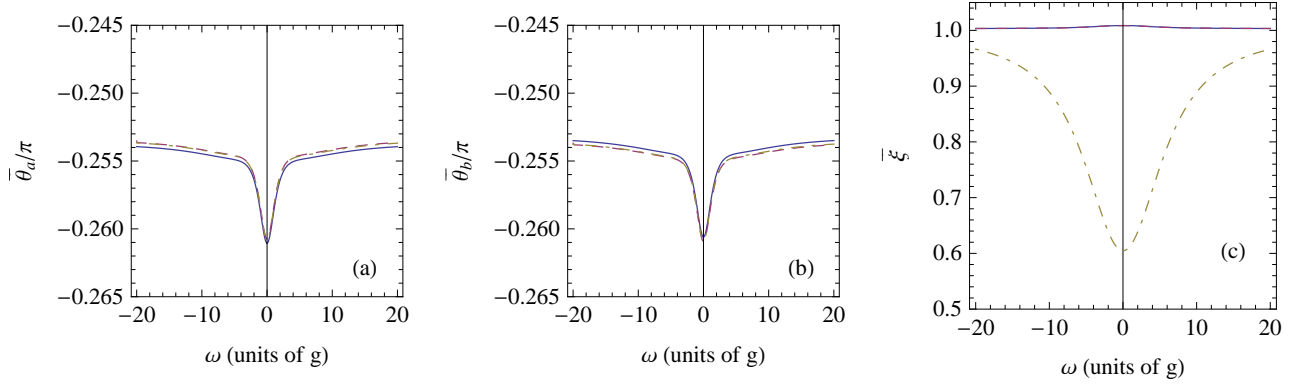}
\caption[]{(Color online) Values of (a) $\theta_a$, (b) $\theta_b$ and (c) $\xi$ that minimize the squeezing spectrum in Fig.~\ref{f1} (a) for each value of $\omega$ (in units of $g$).
}
\label{f2}
\end{figure*}

The Fourier transform of the composite quadrature defined in Eq.~\rp{X:out} is
\begin{eqnarray}
\widetilde X_{\rm out}\al{\theta_a,\theta_b,\xi}(\omega)&=&\frac{\abs{\xi}}{\sqrt{\xi^4+1}}
\nn\\&&\times
\lpq{\abs{\xi}\tilde a_{\rm out}^{(d)}(\omega+\Delta_a)\ee^{\ii \theta_a}+\abs{\xi}\tilde a_{\rm out}^{\dagger(d)}(\omega-\Delta_a)\ee^{-\ii\theta_a}
}\nn\\&&\rpq{
+\frac{1}{\xi}b_{\rm out}^{(d)}(\omega+\Delta_b)\ee^{\ii\theta_b}+\frac{1}{\xi}b_{\rm out}^{\dagger(d)}(\omega-\Delta_b)\ee^{-\ii\theta_b}
}\nn\,.\\
\label{X:out_omega}
\end{eqnarray}
The two-mode squeezing spectrum, is obtained by integrating the corresponding two-frequency correlation function $\Delta X_{\rm out}\al{\theta_a,\theta_b,\xi}(\omega,\omega')=\av{\widetilde X_{\rm out}\al{\theta_a,\theta_b,\xi}(\omega)\ \widetilde X_{\rm out}\al{\theta_a,\theta_b,\xi}(\omega')}$ over a small range of frequency $\omega'$  around $-\omega$~\cite{Mandel}
\begin{eqnarray}
S_\xi\al{\theta_a,\theta_b}(\omega)&=&\lim_{\delta_\omega\to0}\int_{-\omega-\delta_\omega/2}^{-\omega+\delta_\omega/2}\dd\omega'\
\Delta X_{\rm out}\al{\theta_a,\theta_b,\xi}(\omega,\omega').
\nn\\
\end{eqnarray}
When $\Delta_a=-\Delta_b\equiv\Delta\neq 0$, that is the case that we consider in this article (see Fig.~\ref{f1}), the squeezing spectrum can be written in terms of the elements of the matrix~\rp{AA0} as
\begin{eqnarray}\label{Somega}
&& S_\xi\al{\theta_a,\theta_b}(\omega)=\frac{\xi^2}{\xi^4+1}\lpg{
\xi^2\pq{
\pg{\widetilde\AAA_0\pt{\omega+\Delta}}_{1,3}+\pg{\widetilde\AAA_0\pt{\omega-\Delta}}_{3,1}
}
}\nn\\&&
+\frac{1}{\xi^2}\pq{
\pg{\widetilde\AAA_0\pt{\omega+\Delta}}_{4,2}+\pg{\widetilde\AAA_0\pt{\omega-\Delta}}_{2,4}
}
\nn\\&&+\frac{\abs{\xi}}{\xi}
\ee^{\ii(\theta_a+\theta_b)}\pq{
\pg{\widetilde\AAA_0\pt{\omega+\Delta}}_{1,2}+\pg{\widetilde\AAA_0\pt{\omega-\Delta}}_{2,1}}
\nn\\&&+\rpg{ \frac{\abs{\xi}}{\xi}
\ee^{-\ii(\theta_a+\theta_b)}\pq{
\pg{\widetilde\AAA_0\pt{\omega+\Delta}}_{4,3}+\pg{\widetilde\AAA_0\pt{\omega-\Delta}}_{3,4}}
}\ .
\end{eqnarray}
The plots in Fig.~\ref{f1} (a) and (b) have been evaluated using this expression and for the values of the parameters $\theta_a$, $\theta_b$ and $\xi$ that minimize the squeezing spectrum and the entanglement criterion in Eq.~\rp{S:sum}. These values are reported in Fig.~\ref{f2}. They depend on $\omega$, therefore these curves can be experimentally obtained by sampling the squeezing spectra for different values of $\xi$.

\begin{widetext}
The corresponding logarithmic negativity is reported in Fig.~\ref{f1} (c). It has been evaluated using the correlation matrix for the detected field operators at frequency $\omega$, whose elements are
$\pg{\widetilde \AAA\al{\Delta}(\omega,\omega')}_{j,k}=\av{\pg{\tilde{\va}_\Delta^{\rm out}(\omega)}_j\ \pg{\tilde{\va}_\Delta^{\rm out}(\omega')}_k}
$, with $\va^{\rm out}_\Delta(\omega)$ the vector of detected field operators
$$\va^{\rm out}_\Delta(\omega)=\pt{
\frac{\tilde a_{\rm out}\al{d}(\omega+\Delta)+\tilde a_{\rm out}\al{d}(-\omega+\Delta)}{\sqrt{2}},
\frac{\tilde b_{\rm out}\al{d}(\omega-\Delta)+\tilde b_{\rm out}\al{d}(-\omega-\Delta)}{\sqrt{2}},
\frac{\tilde a_{\rm out}^{\dagger(d)}(\omega-\Delta)+\tilde a_{\rm out}^{\dagger(d)}(-\omega-\Delta)}{\sqrt{2}},
\frac{\tilde b_{\rm out}^{\dagger(d)}(\omega+\Delta)+\tilde b_{\rm out}^{\dagger(d)}(-\omega+\Delta)}{\sqrt{2}}
}^T.$$
In particular, the matrix $\widetilde \AAA\al{\Delta}(\omega,\omega')$ is given by
\begin{eqnarray}
\widetilde \AAA\al{\Delta}(\omega,\omega')&=&\av{\tilde{\va}_\Delta^{\rm out}(\omega)\ \tilde{\va}_\Delta^{\rm out}(\omega')^T}
=\delta(\omega+\omega') \frac{1}{2}\pq{\widetilde \AAA_0\al{\Delta}(\omega)+\widetilde \AAA_0\al{\Delta}(-\omega)}\ .
\end{eqnarray}
where $\widetilde \AAA_0\al{\Delta}(\omega)$ can be written in terms of the elements of the matrix~\rp{AA0} as
\begin{eqnarray}
\widetilde A_0\al{\Delta}(\omega)=
\frac{1}{2}\pt{
\begin{array}{cccc}
0&\pg{\widetilde\AAA_0\pt{\omega+\Delta}}_{1,2}&\pg{\widetilde\AAA_0\pt{\omega+\Delta}}_{1,3}&0 \\
\pg{\widetilde\AAA_0\pt{\omega-\Delta}}_{2,1}&0&0&\pg{\widetilde\AAA_0\pt{\omega-\Delta}}_{2,4} \\
\pg{\widetilde\AAA_0\pt{\omega-\Delta}}_{3,1}&0&0&\pg{\widetilde\AAA_0\pt{\omega-\Delta}}_{3,4} \\
0&\pg{\widetilde\AAA_0\pt{\omega+\Delta}}_{4,2}&\pg{\widetilde\AAA_0\pt{\omega+\Delta}}_{4,3}&0
\end{array}
}\ .
\end{eqnarray}
The logarithmic negativity has been evaluated according to the definition of Sec.~\ref{LogNeg} applied to the covariance matrix corresponding to the correlation matrix $\pq{\widetilde \AAA_0\al{\Delta}(\omega)+\widetilde \AAA_0\al{\Delta}(-\omega)}/2$.
\end{widetext}


\begin{thebibliography}{breitestes Label}

\bibitem{Horodecki}
R. Horodecki, P. Horodecki, M. Horodecki, and K. Horodecki,
Rev. Mod. Phys. {\bf 81}, 865 (2009).

\bibitem{EPR}
A. Einstein, B. Podolsky, and N. Rosen, Phys. Rev. {\bf 47}, 777 (1935).

\bibitem{Reid1989}
M. D. Reid, Phys. Rev. A {\bf 40}, 913 (1989).

\bibitem{Reid2009}
M. D. Reid, P. D. Drummond, W. P. Bowen, E. G. Cavalcanti, P. K. Lam, H. A. Bachor, U. L. Andersen, and G. Leuchs, Rev. Mod. Phys. {\bf 81}, 1727 (2009).

\bibitem{Duan}
L.-M. Duan, G. Giedke, J. I. Cirac, and P. Zoller, Phys. Rev. Lett. {\bf 84}, 2722 (2000)

\bibitem{Adesso}
G. Adesso and F. Illuminati,
J. Phys. A: Math. Theor. {\bf 40},  7821 (2007).

\bibitem{QSensing}
W. Wasilewski, K. Jensen, H. Krauter, J. J. Renema, M. V. Balabas, and E. S. Polzik, Phys. Rev. Lett. {\bf 104}, 133601 (2010).

\bibitem{Monras}
A. Monras and F. Illuminati, Phys. Rev. A {\bf 81}, 062326 (2010); 
A. Monras and F. Illuminati, Phys. Rev. A {\bf 83}, 012315 (2011).  

\bibitem{QMetrology}
S. Steinlechner, J. Bauchrowitz, M. Meinders, H. M\"uller-Ebhardt, K. Danzmann, and R. Schnabel, Nature Photonics {\bf 7}, 626 (2013).

\bibitem{vanLoock}
S. L. Braunstein and P. van Loock, Rev. Mod. Phys. {\bf 77}, 513 (2005).

\bibitem{Hammerer}
K. Hammerer, A. S. S\o renson, and E. S. Polzik, Rev. Mod. Phys. {\bf 82}, 1041 (2010).

\bibitem{Lee}
N. Lee, H. Benichi, Y. Takeno, S. Takeda, J. Webb, E. Huntington, A. Furusawa,
Science {\bf 332}, 330 (2011).

\bibitem{Weedbrook}
C. Weedbrook, S. Pirandola, R. Garc\'{\i}a-Patr\'on, N. J. Cerf, T. C. Ralph, J. H. Shapiro, and S. Lloyd,
Rev. Mod. Phys. {\bf 84}, 621 (2012). 

\bibitem{Krauter2013}
H. Krauter, D. Salart, C. A. Muschik, J. M. Petersen, Heng Shen, T. Fernholz, E. S. Polzik, Nature Physics {\bf 9}, 400 (2013).

\bibitem{Brakhane}
S. Brakhane, W. Alt, T. Kampschulte, M. Martinez-Dorantes, R. Reimann, S. Yoon, A. Widera, and D. Meschede,
Phys. Rev. Lett., {\bf 109}, 173601 (2012).

\bibitem{Schleier-Schmidt}
M. H. Schleier-Smith, I. D. Leroux, H. Zhang, M. A. Van Camp, and V. Vuleti{\'c}, Phys. Rev. Lett. {\bf 107}, 143005 (2011).

\bibitem{Ou1992}
Z. Y. Ou, S. F. Pereira, H. J. Kimble, and K. C. Peng , Phys. Rev. Lett. {\bf 68}, 3663 (1992).

\bibitem{ParkinsKimble}
A. S. Parkins and H. J. Kimble, J. Opt. B {\bf 1}, 496 (1999); A. Peng and A. S. Parkins, Phys. Rev. A {\bf 65}, 062323 (2002).

\bibitem{Zippilli2004}
S. Zippilli, G. Morigi, and H. Ritsch, Phys. Rev. Lett. {93}, 123002 (2004).

\bibitem{Vogel}
W. Vogel and D.-G. Welsch,
Phys. Rev. Lett. {\bf 54}, 1802 (1985); P. Gr\"unwald and W. Vogel,
Phys. Rev. Lett., {\bf 104}, 233602 (2010).

\bibitem{FernandezVidal}
S. Fernandez-Vidal, S. Zippilli, and G. Morigi, Phys. Rev. A {\bf 76}, 053829 (2007).

\bibitem{HabibianPRA2011}
H. Habibian, S. Zippilli, and G. Morigi, Phys. Rev. A \textbf{84}, 033829 (2011).

\bibitem{HolsteinPrimakoff}
T. Holstein and H. Primakoff, Phys. Rev. \textbf{58}, 1098-1113 (1940).

\bibitem{Vidal}
G. Vidal and R. F. Werner, Phys. Rev. A \textbf{65}, 032314 (2002).

\bibitem{Plenio}
M. B. Plenio, Phys. Rev. Lett. {\bf 95}, 090503 (2005).

\bibitem{Reid1990}
P. D. Drummond and M. D. Reid,
 Phys. Rev. A, {\bf 41}, 3930 (1990).

\bibitem{Reid2004}
K. Dechoum, P. D. Drummond, S. Chaturvedi, and M. D. Reid, Phys. Rev. A {\bf 70}, 053807 (2004).

\bibitem{MilburnWalls}
D. F. Walls and G. J. Milburn, \textit{Quantum Optics} (Springer, Berlin, 1994).

\bibitem{Vitali2006}
D. Vitali, G. Morigi, and J. Eschner, Phys. Rev. A {\bf 74}, 053814 (2006).

\bibitem{PeresHorodeckis}
A. Peres, \prl \textbf{77}, 1413 (1996);
M. Horodecki, P. Horodecki, and R. Horodecki, Phys. Lett. A \textbf{223}, 1 (1996).

\bibitem{Werner}
R. F. Werner and M. M. Wolf, \prl \textbf{86}, 3658 (2001).

\bibitem{Simon}
R. Simon, Phys. Rev. Lett. {\bf 84}, 2726 (2000).

\bibitem{Vuletic}
M. H. Schleier-Smith, I. D. Leroux, and V. Vuleti\'c,
Phys. Rev. A, {\bf 81}, 021804 (2010);
M. H. Schleier-Smith, I. D. Leroux, and V. Vuleti\'c,
Phys. Rev. Lett., {\bf 104}, 073604, (2010);
I. D. Leroux, M. H. Schleier-Smith, and V. Vuleti\'c,
Phys. Rev. Lett., {\bf  104}, 073602, (2010).
I. D. Leroux, M. H. Schleier-Smith, and V. Vuleti\'c,
Phys. Rev. Lett. {\bf 104}, 250801 (2010).
H. Zhang, R. McConnell, S. \'Cuk, Q. Lin, M. H. Schleier-Smith, I. D. Leroux, and V. Vuleti\'c,
Phys. Rev. Lett. {\bf 109}, 133603 (2012).

\bibitem{Pirandola03}
S. Pirandola, S. Mancini, D. Vitali, and P. Tombesi, Phys. Rev. A \textbf{68},
062317 (2003).

\bibitem{Pirandola06}
S. Pirandola and S. Mancini, Laser Phys. {\bf 16}, 1418 (2006).

\bibitem{Dellanno2007}
F. Dell'Anno, S. De Siena, L. Albano, and F. Illuminati,
Phys. Rev. A {\bf 76}, 022301 (2007). 

\bibitem{Rebic2010}
S. Rebic, S. Mancini, G. Morigi, and D. Vitali,
J. Opt. Soc. Am. B {\bf 27}, A198 (2010).

\bibitem{Dellanno2013}
F. Dell'Anno, D. Buono, G. Nocerino, A. Porzio, S. Solimeno, S. De Siena, and F. Illuminati,
Phys. Rev. A {\bf 88}, 043818 (2013).

\bibitem{Genoni}
M. G. Genoni, M. G. A. Paris, and K. Banaszek,
Phys. Rev. A {\bf 76}, 042327 (2007); 
M. G. Genoni, M. G. A. Paris, and K. Banaszek,
Phys. Rev. A {\bf 78}, 060303 (2008); 
M. G. Genoni and M. G. A. Paris,
Phys. Rev. A {\bf 82}, 052341 (2010);
M. G. Genoni, M. L. Palma, T. Tufarelli, S. Olivares, M. S. Kim, and M. G. A. Paris,
Phys. Rev. A {\bf 87}, 062104 (2013).

\bibitem{Dellanno2010}     
F. Dell'Anno, S. De Siena, and F. Illuminati,
Phys. Rev. A {\bf 81}, 012333 (2010). 

\bibitem{Muschik}
C. A. Muschik, E. S. Polzik, and J. I. Cirac,
Phys. Rev. A, {\bf 83}, 052312 (2011).

\bibitem{Krauter}
H. Krauter, C. A. Muschik, K. Jensen, W. Wasilewski, J. M. Petersen, J. I. Cirac, and E. S. Polzik,
Phys. Rev. Lett., {\bf 107}, 080503 (2011).


\bibitem{MorigiPRL2006}
G. Morigi, J. Eschner, S. Mancini, and D. Vitali,
Phys. Rev. Lett. {\bf 96}, 023601 (2006).

\bibitem{Campbell2010}
G. Adesso, S. Campbell, F. Illuminati, and M. Paternostro,
Phys. Rev. Lett. {\bf 104}, 240501 (2010). 

\bibitem{Zippilli2013}
S. Zippilli, M. Paternostro, G. Adesso, and F. Illuminati,
Phys. Rev. Lett. {\bf 110}, 040503 (2013). 


\bibitem{Mandel}
L. Mandel and E. Wolf, \textit{Optical coherence and quantum optics} (Cambridge University Press, 1995).




\end{thebibliography}
\end{document}